\documentclass[11pt,a4paper]{article}

\usepackage{jheppub}
\usepackage{latexsym}
\usepackage{multirow}
\usepackage{color}
\usepackage[usenames,dvipsnames,table]{xcolor}
\usepackage{graphicx}
\usepackage{epsfig}  
\usepackage{epsf}    
\usepackage{dcolumn}
\usepackage{bm}
\usepackage{dcolumn}
\usepackage{textcomp}
\usepackage{float}
\usepackage{subfig}

\usepackage[toc,page]{appendix}

\usepackage{amsmath}
\usepackage[mathlines]{lineno}

\usepackage{hypcap}
\usepackage[]{hyperref}
\usepackage{makecell}
\usepackage{color}
\usepackage{pifont}
\usepackage{appendix}
\usepackage{multirow,bigdelim}  
\usepackage{lineno}
\usepackage[normalem]{ulem}
\hypersetup{
	bookmarks=true,         
	unicode=false,          
	pdftoolbar=true,        
	pdfmenubar=true,        
	pdffitwindow=true,     
	pdfstartview={FitH},    
	pdfsubject={Neutrino Oscillations Phenomenology},   
	pdfnewwindow=true,      
	pdfcreator={RevTeX},
	colorlinks=true,       
	linkcolor=red,          
	citecolor=blue,        
	filecolor=black,      
	urlcolor=blue,           
}

\newcommand{\be}{\begin{equation}}
\newcommand{\ee}{\end{equation}}
\newcommand{\ba}{\begin{eqnarray}}
\newcommand{\ea}{\end{eqnarray}}

\newcommand{\capdef}{}
\newcommand{\mycaption}[2][\capdef]{\renewcommand{\capdef}{#2}
	\caption[#1]{{\footnotesize #2}}}
\makeatletter
\renewcommand{\fnum@table}{\textbf{\tablename~\thetable}}
\renewcommand{\fnum@figure}{\textbf{\figurename~\thefigure}}
\makeatother

\preprint{IP/BBSR/2024-03, TIFR/TH/24-05}

\title{Constraining the core radius and density jumps inside Earth using atmospheric neutrino oscillations}

\author[a,b]{Anuj Kumar Upadhyay,}
\author[b,c,d]{Anil Kumar,}
\author[b,d,e]{Sanjib Kumar Agarwalla,}
\author[f]{Amol Dighe} 

\affiliation[a]{Department of Physics, Aligarh Muslim University, Aligarh-202002, India }
\affiliation[b]{Institute of Physics, Sachivalaya Marg, Sainik School Post, Bhubaneswar 751005, India}
\affiliation[c]{Applied Nuclear Physics Division, Saha Institute of Nuclear Physics, Block AF, Sector 1, Bidhannagar, Kolkata 700064, India}
\affiliation[d]{Homi Bhabha National Institute, Anushakti Nagar, Mumbai 400094, India}
\affiliation[e]{Department of Physics \& Wisconsin IceCube Particle Astrophysics Center, University of Wisconsin, Madison, WI 53706, U.S.A}
\affiliation[f]{Department of Theoretical Physics, Tata Institute of Fundamental Research, Homi Bhabha Road, Colaba, Mumbai 400005, India}

\emailAdd{anuju@iopb.res.in  (ORCID: 0000-0003-1957-2626)}
\emailAdd{anil.k@iopb.res.in (ORCID: 0000-0002-8367-8401)}
\emailAdd{sanjib@iopb.res.in (ORCID: 0000-0002-9714-8866)}
\emailAdd{amol@theory.tifr.res.in (ORCID: 0000-0001-6639-0951)}

\abstract{Atmospheric neutrinos probe the interior of Earth using weak interactions, and provide information complementary to that of gravitational and seismic measurements. While passing through Earth, multi-GeV neutrinos encounter matter effects due to the coherent forward scattering with ambient electrons, which alter the neutrino oscillation probabilities. These matter effects depend upon the density distribution of electrons inside Earth, and hence, can be used to determine the internal structure of Earth. In this work, we employ a five-layered model of Earth where the layer densities and radii are modified, keeping the mass and moment of inertia of Earth unchanged and respecting the hydrostatic equilibrium condition. We use the proposed INO-ICAL detector as an example of an atmospheric neutrino experiment that can distinguish between neutrinos and antineutrinos efficiently in the multi-GeV energy range. Our analyses demonstrate that such an experiment can simultaneously constrain density jumps inside Earth and locate the core-mantle boundary. The charge identification (CID) capability of the ICAL detector would play a crucial role in obtaining these correlated constraints. An ICAL-like detector without CID capability would also be able to perform this task, albeit with a reduced sensitivity.}

\arxivnumber{2405.04986}

\begin{document}
\maketitle
\flushbottom

\section{Introduction} 
\label{sec:introduction}

The knowledge of the internal structure of Earth is important not only for understanding the effects of its dynamics on the surface, but also as an input for reconstructing the formation and geological evolution of Earth. The direct probing of the internal structure and chemical composition of Earth is impractical. Moreover, given the large temperature, pressure, and extreme environment within the interior of Earth, it is currently not feasible to replicate the same conditions in the lab. Therefore, most of the information about the internal structure of Earth can only be derived from indirect methods, mainly gravitational measurements~\cite{Ries:1992,Luzum:2011,Rosi:2014kva,astro_almanac,Williams:1994,Chen:2014} and seismic studies~\cite{Gutenberg:1914,robertson1966interior,Volgyesi:1982,Loper:1995,Alfe:2007,Stacey_Davis:2008,McDonough:2022,Thorne:2022,Hirose:2022,McDonough_MMTE:2023}.

Gravitational measurements provide the values of the mass of Earth, $M_E = (5.9722\,\pm\,0.0006) \times 10^{24}$ kg~\cite{Ries:1992,Williams:1994,Luzum:2011,Rosi:2014kva,astro_almanac}, and its moment of inertia, $I_E = (8.01736\,\pm\, 0.00097) \times 10^{37}$ kg$\cdot$m\textsuperscript{2}~\cite{Williams:1994,Chen:2014}. These measurements already indicate that the density of Earth cannot be uniform; rather, the density closer to the center of Earth should be larger. This may be inferred from the following. A uniform sphere of mass $M_E$ and radius $R_E = 6371$ km would have a moment of inertia $I_E = (2/5) M_E R_E^2 = 9.7 \times 10^{37}$ kg$\cdot$m\textsuperscript{2}, which is significantly higher than the experimentally measured value of $I_E$. Also, the average density of Earth obtained from $M_E$ and $R_E$, i.e., $\rho_\text{avg} \sim5.5$ g/cm$^3$, is larger than the density of rocks at the surface, $\rho \sim 2.8$ g/cm$^3$. Both of these observations are consistent with a high-density region present deep inside Earth.

The details of this increase in density as we go deeper inside Earth may be discerned from seismological measurements. Seismic waves originate from the epicenter of an earthquake, which is typically located at depths of up to 700 km~\cite{Volgyesi:1982,Stacey_Davis:2008}. These are mechanical waves that propagate through different regions inside Earth, and their velocities modify depending upon the properties of the media they traverse. They can reflect or refract when they encounter changes in densities, pressures, and physical properties of the medium. Seismic studies indicate that Earth has a layered structure in the form of concentric spherical shells. The interior of Earth can be broadly divided into two major layers; the mantle and the core, each of which contains several numbers of sub-layers. The sub-layers of the mantle can be taken to be the outer and inner mantle, while the sub-layers of the core can be taken to be the outer and inner core. Additionally, there is an outermost layer near Earth’s surface known as the crust. The density increases as we go deeper inside towards the inner core. This increase in density is not always gradual; there are sharp density changes in the transition regions between any two of the layers. The sharpest such transition is observed while going from the inner mantle to the outer core; this core-mantle boundary (CMB) is located around $R_\text{CMB} = 3483\pm5$ km~\cite{Gutenberg:1914,young1987core,Masters1995,McDonough:2003,McDonough:2024}. Based on seismic studies and the expected composition of Earth's core, the density of the core is not expected to be higher than 14-15 g/cm$^3$~\cite{McDonough:2024}. However, the aim of our neutrino tomography work is to quantify what the neutrino data by itself can tell us about the internal structure of Earth, with minimal inputs from seismological data. Therefore, in our work, we explore a wide range of layer densities for analysing the interior of Earth, while keeping the overall layer structure consistent with seismological data.

The density distribution inside Earth is obtained by the nonlinear inversion of velocity distributions of the seismic waves. This indirect technique is used to develop Earth density models~\cite{Gilbert-Dziewonski1975,Dziewonski-Hales-Lapwood1975,Kennett-Engdahl1991,Kennett-etal1995,Cammarano-etal-2005,Kustowski-etal-2008}; one of the well-known models is the Preliminary Reference Earth Model (PREM)~\cite{Dziewonski:1981xy}. It presents the density of any layer as a one-dimensional function\footnote{The seismic measurements suggest that the deviation from the spherical symmetry near the CMB is very minimal. For instance, the topography of the outer core surfaces varies within 3 km, while the ellipticity at the CMB is approximately $2.5 \times 10^{-3}$~\cite{McDonough2017}. In recent years, several new three-dimensional Earth models have been developed, including FWEA18~\cite{FWEA18:2018}, Shen-Ritzwoller (S-R)~\cite{sr:2016}, CRUST1~\cite{CRUST1:2013}, SAW642AN~\cite{SAW642AN:2000}. As far as neutrinos are concerned, they are not expected to be significantly sensitive to these detailed features.} of the radial distance of the layer from the centre of Earth. The seismic wave velocities depend upon temperature, pressure, composition, and elastic properties of the layers of Earth, which give rise to uncertainties in the determination of the density distribution. The density of the mantle is uncertain by about 5\%, whereas the core is the most uncertain region of Earth~\cite{Bolt:1991,kennett:1998,Masters:2003}.
	
Even though the huge amount of data from seismic studies and gravitational measurements has significantly improved our understanding of the internal structure of Earth, we still have many open questions. For example, the mass, the radius, and the chemical composition of the core, as well as the density jump while going from the outer core to the inner core, still have uncertainties~\cite{McDonough:2024}. The uncertainties in the radius of the inner and outer core are approximately 10 km and 5 km, respectively~\cite{Masters1995}. Ref.~\cite{McDonough:2024} also presents a comparison of the density jump at the inner-outer core boundary as predicted by different seismological models. As far as the uncertainties in chemical composition are concerned, the abundances of Fe, Ni, Co, W, Mo, and other elements in the core have about 15\% statistical uncertainty, whereas the uncertainties in O, Si, and S content of the core are estimated to range from 25\% to 50\%~\cite{McDonough:2024}. The fraction of a light element like H in the core is also unknown~\cite{Williams:2001,Hirose:2021,Hirose:2022}. The measurement of power produced during radioactive decays inside the core and the mantle could help us understand the thermal dynamics~\cite{Araki:2005qa,Borexino:2015ucj,Bellini:2021sow,KamLAND:2022vbm}. Complementary and non-traditional probes such as geoneutrino detection~\cite{Araki:2005qa,Fiorentini:2007te,Bellini:2013wsa,McDonough2015,Michael2017,Bellini:2021sow}, neutrino absorption~\cite{Winter:2006vg,Gonzalez-Garcia:2007wfs,Donini:2018tsg}, and neutrino oscillations~\cite{Winter:2015zwx} could further improve our understanding about the interior of Earth. The measurement of the geoneutrino flux has already provided new insights into the amount of Earth's radiogenic heat production and the composition of the continental crust and mantle~\cite{McDonough:2022a}. On the other hand, neutrino absorption and oscillation tomography hold the potential to probe the deep Earth's interior in the future.

Neutrinos are an appealing tool for providing a complementary pathway to learn about the interior of Earth via their weak interactions with matter, which is complementary to the electromagnetic interactions of seismic studies and gravitational interactions.  The idea of exploiting the attenuation of neutrinos at energies above a few TeV~\cite{Gandhi:1995tf,IceCube:2017roe} to learn about the Earth's interior was first suggested in Refs.~\cite{Placci:1973,Volkova:1974xa}. The detailed studies using various neutrino sources, such as man-made neutrinos~\cite{Placci:1973,Volkova:1974xa,Nedyalkov:1981,Nedyalkov:1981pp,Nedyalkov:1981yy,Nedialkov:1983,Krastev:1983,DeRujula:1983ya,Wilson:1983an,Askarian:1984xrv,Volkova:1985zc, Tsarev:1985yub,Borisov:1986sm,Tsarev:1986xg,Borisov:1989kh,Winter:2006vg}, extraterrestrial neutrinos~\cite{Wilson:1983an,Kuo:1995,Crawford:1995,Jain:1999kp,Reynoso:2004dt}, and atmospheric neutrinos~\cite{Gonzalez-Garcia:2007wfs,Donini:2018tsg,Borriello:2009ad,Takeuchi:2010,Romero:2011zzb}, have also been carried out. On the other hand, the discovery of a non-zero value of reactor mixing angle $\theta_{13}$, and precise measurement of other neutrino oscillation parameters~\cite{Capozzi:2025wyn,NuFIT6.0,Esteban:2024eli,deSalas:2020pgw} have provided a complementary probe to the Earth's interior using matter effects in neutrino oscillations at energies of about a few GeV. This possibility is called ``neutrino oscillation tomography'', and has been studied using man-made neutrino beams~\cite{Ermilova:1986ph,Nicolaidis:1987fe,Ermilova:1988pw,Nicolaidis:1990jm,Ohlsson:2001ck,Ohlsson:2001fy,Winter:2005we,Minakata:2006am,Gandhi:2006gu,Tang:2011wn,Arguelles:2012nw}, solar neutrinos~\cite{Ioannisian:2002yj,Ioannisian:2004jk,Akhmedov:2005yt,Ioannisian:2015qwa,Ioannisian:2017chl,Ioannisian:2017dkx,Bakhti:2020tcj}, supernova neutrinos~\cite{Lindner:2002wm,Akhmedov:2005yt,Hajjar:2023knk}, and atmospheric neutrinos~\cite{Agarwalla:2012uj,IceCube-PINGU:2014okk,Rott:2015kwa,Winter:2015zwx,Bourret:2017tkw,Bourret:2019wme,Bourret:2020zwg,DOlivo:2020ssf,Kumar:2021faw,Maderer:2021aeb,Denton:2021rgt,Kelly:2021jfs,Capozzi:2021hkl,DOlivoSaez:2022vdl,Maderer:2022toi,Upadhyay:2022jfd,Upadhyay:2021kzf}.

Atmospheric neutrinos are ideally suited for performing Earth tomography because they have energies covering the multi-GeV range where Earth matter effects are significant, and a wide range of baselines starting from $\sim15$ km to 12757 km. Neutrinos passing through the mantle predominantly experience flavor transitions enhanced by the Mikheyev-Smirnov-Wolfenstein (MSW) resonance~\cite{Wolfenstein:1977ue,Mikheev:1986gs,Mikheev:1986wj}, particularly in the energy range of 6 to 10 GeV. The core-passing neutrinos with energies of $3-6$ GeV, in addition to the MSW resonance corresponding to the core density, experience another kind of resonance, which is known as the parametric resonance (PR)~\cite{Ermilova:1986,Akhmedov:1988kd,Krastev:1989,Akhmedov:1998ui,Akhmedov:1998xq} or neutrino oscillation length resonance (NOLR)~\cite{Petcov:1998su,Chizhov:1998ug,Petcov:1998sg,Chizhov:1999az,Chizhov:1999he}. This additional enhancement arises due to the low, high, and then again low densities encountered by these neutrinos as they cross the core-mantle boundaries twice. The efficient observations of these matter effect resonances could help us in probing the physical and chemical properties of the core and the mantle regions inside Earth. Using matter effects in atmospheric neutrino oscillations, numerous sensitivity studies has been performed recently for current and future experiments like the IceCube~\cite{IceCube:2016zyt}, DeepCore~\cite{IceCube:2011ucd, IceCube:2016zyt}, Precision IceCube Next Generation Upgrade (PINGU)~\cite{IceCube-PINGU:2014okk}, Oscillation Research with Cosmics in the Abyss (ORCA)~\cite{KM3Net:2016zxf}, Deep Underground Neutrino Experiment (DUNE)~\cite{DUNE:2021tad}, Hyper-Kamiokande (Hyper-K)~\cite{Hyper-Kamiokande:2018ofw}, and Iron Calorimeter (ICAL)~\cite{ICAL:2015stm} detector. These studies include validating the core-mantle boundary (CMB) using ICAL~\cite{Kumar:2021faw,Anil_MMTE:2023}, determining the position of the core-mantle boundary using DUNE~\cite{Denton:2021rgt} and ICAL~\cite{Upadhyay:2022jfd,Anil_MMTE:2023,Anuj_MMTE:2023}, determining the average densities of the core and the mantle using ORCA~\cite{Winter:2015zwx,Maderer:2021aeb,Capozzi:2021hkl} and DUNE~\cite{Kelly:2021jfs}, probing the possible presence of dark matter inside Earth using ICAL~\cite{Upadhyay:2021kzf,Anil_MMTE:2023}, and exploring the chemical composition of the Earth's core using PINGU~\cite{IceCube-PINGU:2014okk}, Hyper-K and IceCube~\cite{Rott:2015kwa}, and ORCA~\cite{Bourret:2017tkw,Bourret:2019wme,Bourret:2020zwg,Maderer:2021aeb,Maderer:2022toi,DOlivoSaez:2022vdl}. The prospects of establishing the Earth's matter effects, validating the non-homogeneous density profile inside Earth, and measuring the mass of Earth and correlated densities of its different layers using oscillations of the weakly interacting neutrino at GeV energies in the context of the IceCube DeepCore are presented in Ref.~\cite{Chattopadhyay:2025cgt}.

Detectors that can distinguish neutrinos from antineutrinos have a special role to play in this exploration, as we shall present later in this paper (see appendix~\ref{app:sen_regions} for a detailed discussion). Therefore, we focus on the proposed magnetized Iron Calorimeter (ICAL) detector at the India-based Neutrino Observatory (INO)~\cite{ICAL:2015stm}, which is optimized to detect multi-GeV atmospheric neutrinos and antineutrinos separately.  Owing to its excellent angular resolution, it would be able to observe neutrinos passing through the core and mantle. The good energy resolution would enable ICAL to efficiently observe the MSW resonance and the PR/NOLR of matter effects. These resonances depend on the matter densities of the layers inside Earth and the location of layer boundaries. Thanks to these features, ICAL would be sensitive to the heights of density jumps at the layer boundaries and the location of the CMB. In Ref.~\cite{Kumar:2021faw}, some of the present authors had calculated the expected sensitivity of the ICAL detector to determine the presence of a core-mantle boundary with a given density jump and a given position of the CMB. As a follow-up work in Ref.~\cite{Upadhyay:2022jfd}, we evaluated the expected sensitivity of the ICAL detector to locate the position of the CMB, assuming various cases. 

In the present work, we quantify the sensitivity of the ICAL detector to measure the density jumps at various boundaries and the location of the CMB radius simultaneously. We take into account the constraints from the mass of Earth, the moment of inertia of Earth, and the hydrostatic equilibrium condition of increasing density with decreasing radius. We demonstrate that a five-layered model is sufficient to incorporate these constraints and perform a 2D analysis in the parameter space of the CMB radius and one of the density jumps. We also present the expected sensitivities for individual density jumps when the CMB radius is fixed. We present the expected sensitivity for both with and without the charge identification (CID) capability of the ICAL detector. It is observed that the CID capability of the ICAL detector plays an important role in constraining the allowed parameter space. The sensitivity without CID would correspond to that of a generic detector with identical properties as ICAL, except for its inability to distinguish between neutrinos and antineutrinos.

In section~\ref{sec:density_jump_variation}, we describe a five-layered density profile of Earth used in the present analysis and the procedure to modify the density jumps at layer boundaries and the location of the CMB, while respecting the constraints such as the total mass and the moment of inertia of Earth. Section~\ref{sec:density_jump_oscillation} presents the effects of these modifications on the neutrino oscillation probabilities. In section~\ref{sec:event_genration}, we describe the procedure to simulate the reconstructed events at the ICAL detector and the effects of modified density jumps and the CMB location on reconstructed event distributions. Details regarding numerical analysis are mentioned in section~\ref{sec:analysis_metho}. We present our results in section~\ref{sec:results}, where we evaluate the expected sensitivity of ICAL to measure the density jumps at the layer boundaries and the CMB location. We also show the impact of different true values of the mixing angle $\theta_{23}$ on our sensitivities. Finally, we summarize and conclude in section~\ref{sec:conclusion}. In appendix~\ref{app:sen_regions}, we show how the CID capability of ICAL is important for enhancing sensitivity to matter effects. Appendix~\ref{app:sens_rcmb_measurement} presents the sensitivity for determining the CMB location using both three-layered and five-layered Earth density profiles.
  
\section{Neutrino oscillations with density jumps and location of CMB}

In this section, we explore the effects of variations in the correlated density jumps and the location of the core-mantle boundary ($R_\text{CMB}$) on the oscillation probabilities of atmospheric neutrinos. These variations can modify the matter effects of Earth, which in turn can alter the neutrino oscillation probabilities. Now, we describe the model of density distribution inside Earth, which we have used to perform the present analysis.

\subsection{Five-layered Earth density models with modified density jumps and location of CMB}
\label{sec:density_jump_variation}

In this analysis, we consider a five-layered density profile of Earth\footnote{The rationale for adopting a five-layered profile will be clear later in this section.}, where each layer has a constant density. This density profile is guided by the PREM~\cite{Dziewonski:1981xy} profile. The five distinct layers are the inner core (IC), the outer core (OC), the inner mantle (IM), the outer mantle (OM), and the crust, with respective densities denoted by $\rho_\text{IC}$,  $\rho_\text{OC}$, $\rho_\text{IM}$, $\rho_\text{OM}$, $\rho_\text{crust}$. Each layer's density is taken as the average value calculated from the PREM profile. In this five-layered model, four significant density jumps occur at the following boundaries: 
\begin{itemize}
	\item the inner core - outer core boundary, with radius $R_\text{IC}$ and density jump $\Delta\rho_\text{IC-OC} = \rho_\text{IC} - \rho_\text{OC}$,
	\item the outer core - inner mantle boundary, also known as the CMB, with radius $R_\text{CMB}$ (or $R_\text{OC}$) and density jump $\Delta\rho_\text{CMB} = \rho_\text{OC} - \rho_\text{IM}$,
	\item the inner mantle - outer mantle boundary, with radius $R_\text{IM}$ and density jump $\Delta\rho_\text{IM-OM} = \rho_\text{IM} - \rho_\text{OM}$,
	\item the outer mantle - crust boundary, with radius $R_\text{OM}$ and density jump $\Delta\rho_\text{OM-crust} = \rho_\text{OM} - \rho_\text{crust}$.
\end{itemize}
 The outer radius of the crust is bounded by the radius of Earth $R_\text{E}$. The standard values of the density and radius for each of these layers are given in the first row of Table~\ref{tab:density_jump_variation}.

\begin{table}
	\centering
	\scriptsize
	\bgroup
	\def\arraystretch{1.2}%
	\begin{tabular}{|c|c c c c c|c c c c c|} 
		\hline
		\multirow{2}{*}{Five-layered profile} & \multicolumn{5}{c|}{Radii of layers (km)} & \multicolumn{5}{c|}{ Layer densities $\rho$ (g/$\text{cm}^3$) } \\
		\cline{2-11}
		& $R_\text{IC}$ & $R_\text{OC}$ & $R_\text{IM}$ & $R_\text{OM}$ & $R_\text{crust}$ & $\rho_\text{IC}$ & $\rho_\text{OC}$ & $\rho_\text{IM}$ & $\rho_\text{OM}$ & $\rho_\text{crust}$\\
		\hline\hline
		\textbf{Standard} & 1221.5 & 3480 & 5701 & 6346.6 & 6371 & 12.8936 & 10.9 & 4.9035 & 3.6046 & 2.5204 \\
		\hline
		\textbf{1D-modifications} &  & & & &  & & &  & & \\
		Smaller jump (SJ) & 1221.5 & 3480 & 5701 & 6346.6 & 6371 & 11.1193 & 9.4 & 5.8785 & 2.5267 & 2.5204 \\ 
		Larger jump (LJ)  & 1221.5 & 3480 & 5701 & 6346.6 & 6371 & 13.9582 & 11.8 & 4.3193 & 4.2504 & 2.5204 \\ 
		\hline
		\textbf{2D-modifications} &  & & & & & &  & & &  \\
		Smaller jump at & 1221.5 & 3980 & 5701 & 6346.6 & 6371 & 11.1193 & 9.4 & 4.7338 & 3.4698 & 2.5204 \\ 
		larger core (SJLC) & & & & & & &  & & & \\  
		Larger jump at & 1221.5 & 2980 & 5701 & 6346.6 & 6371 & 13.9582 & 11.8 & 5.6309 & 2.9567 & 2.5204 \\
		smaller core (LJSC) & & & & &  & &  & & & \\ 
		\hline
	\end{tabular}
	\egroup
	\mycaption{The outer radii and densities of the layers in the five-layered profile of Earth for the representative choices shown in Fig.~\ref{fig:dend_profile}. For the 1D-modifications, only $\Delta\rho_\text{CMB}$ is modified, while $R_\text{CMB}$ is kept fixed at its standard value. The smaller jump (SJ) scenario corresponds to $\Delta\rho_\text{CMB}$ of 3.5215 g/cm$^3$, whereas the larger jump (LJ) scenario corresponds to $\Delta\rho_\text{CMB}$ of 7.4807 g/cm$^3$. For the 2D-modifications, both $\Delta\rho_\text{CMB}$ and $R_\text{CMB}$ are modified simultaneously. The smaller jump at a larger core (SJLC) scenario represents $\Delta\rho_\text{CMB}$ of 4.6662 g/cm$^3$ for a given shift of $\Delta R_\text{CMB} = + \, 500$ km, whereas the larger jump at a smaller core (LJSC) scenario represents $\Delta\rho_\text{CMB}$ of 6.1691 g/cm$^3$ for $\Delta R_\text{CMB} = - \, 500$ km. The total mass and moment of inertia of Earth are kept unchanged during all these modifications.}
	\label{tab:density_jump_variation}
\end{table}

In the five-layered model, the total mass ($M_\text{E}$) and moment of inertia ($I_\text{E}$) of Earth are given by

\noindent
\begin{align}
M_\text{E} &= \frac{4\pi}{3}\biggl[\rho_\text{IC}R^3_\text{IC} + \rho_\text{OC}\left(R^3_\text{CMB} - R^3_\text{IC}\right) + \rho_\text{IM}\left(R^3_\text{IM} - R^3_\text{CMB}\right) + \rho_\text{OM}\left(R^3_\text{OM} - R^3_\text{IM}\right) \nonumber \\ 
& + \rho_\text{crust}\left(R^3_\text{E} - R^3_\text{OM}\right) \biggr]\,,\label{eq:mass}\\
I_\text{E} &= \frac{8\pi}{15}\biggl[\rho_\text{IC}R^5_\text{IC} + \rho_\text{OC}\left(R^5_\text{CMB} - R^5_\text{IC}\right) + \rho_\text{IM}\left(R^5_\text{IM} - R^5_\text{CMB}\right) + \rho_\text{OM}\left(R^5_\text{OM} - R^5_\text{IM}\right) \nonumber \\
& + \rho_\text{crust}\left(R^5_\text{E} - R^5_\text{OM}\right) \biggr] \label{eq:moment_of_inertia}\,.
\end{align}
$M_\text{E}$ and $I_\text{E}$ can also be written in terms of the density jumps at the layer boundaries ($ \Delta \rho_\text{IC-OC}$, $\Delta \rho_\text{CMB}$, $ \Delta \rho_\text{IM-OM}$, and $\Delta \rho_\text{OM-crust}$):

\noindent 
\begin{align}
M_\text{E} &= \frac{4\pi}{3}\biggl[ \Delta \rho_\text{IC-OC} R^3_\text{IC} + \Delta \rho_\text{CMB}R^3_\text{CMB} + \Delta \rho_\text{IM-OM}R^3_\text{IM} + \Delta \rho_\text{OM-crust}R^3_\text{OM} + \rho_\text{crust}R^3_\text{E} \biggr]\,,\label{eq:mass_delta_rho}\\
I_\text{E} &= \frac{8\pi}{15}\biggl[ \Delta \rho_\text{IC-OC} R^5_\text{IC} + \Delta \rho_\text{CMB}R^5_\text{CMB} + \Delta \rho_\text{IM-OM}R^5_\text{IM} + \Delta \rho_\text{OM-crust}R^5_\text{OM} + \rho_\text{crust}R^5_\text{E} \biggr] \label{eq:moment_of_inertia_delta_rho}\,.
\end{align}
Equations~\ref{eq:mass_delta_rho} and \ref{eq:moment_of_inertia_delta_rho} have in total ten parameters corresponding to the four density jumps, density of crust $\rho_\text{crust}$, and the five radii at the layer boundaries. We consider the following constraints:

\begin{itemize}
	\item The total mass $M_E = 5.973\times10^{24}$ kg and moment of inertia $I_E=8.127\times 10^{37}$ kg$\cdot$m$^2$ of Earth are taken to be fixed.
	\item Since the density of the crust is very well-known, $\rho_\text{crust}$ is taken to be fixed at $\rho_\text{crust} = 2.5204$ g/cm$^3$.
	\item The radius of Earth has been taken to be $R_E = 6371$ km. The radii of the inner mantle and the outer mantle are fixed at $R_\text{IM} = 5701$ km and $R_\text{OM} = $ 6346.6 km, respectively. 
\end{itemize}
Note that the above constraints are written in the decreasing order of strictness. Thus, we have a total of 6 constraints and we are left with 4 free parameters, all related to the core, which we choose to be $R_\text{CMB}$, $\Delta\rho_\text{CMB}$, $R_\text{IC}$, and $\rho_\text{IC}/\rho_\text{OC}$. Since we would like to focus on the sensitivity to $R_\text{CMB}$ and $\Delta\rho_\text{CMB}$ in this paper, we keep $R_\text{IC} = 1221.5$ km and  $\rho_\text{IC}/\rho_\text{OC} = $ 1.183 fixed in our analysis. These values are close to the values in the standard PREM profile.

All the above constraints have the form of equalities. We also impose an additional constraint that has the form of an inequality. This is the hydrostatic equilibrium condition where the density of the inner layer is always greater than that of the outer layer ($\rho_\text{inner layer} > $ $\rho_\text{outer layer}$). This constraint does not affect the number of degrees of freedom, which stays at two, viz. $R_\text{CMB}$ and $\Delta\rho_\text{CMB}$. The implications of the constraints from Earth's total mass, moment of inertia, and hydrostatic equilibrium condition on neutrino tomography studies have been discussed in detail in Ref.~\cite{Petcov:2024icq}.

The procedure we adopt for incorporating the above constraints involves solving  Eqs.~\ref{eq:mass} and \ref{eq:moment_of_inertia} in terms of the two independent parameters\footnote{Note that any one of the other density jumps, like $\Delta\rho_\text{IC-OC}$, $\Delta\rho_\text{IM-OM}$, or $\Delta\rho_\text{OM-crust}$, can also be chosen as the second independent parameter. In our result section, we will also interpret our sensitivities in terms of these density jumps.} $R_\text{CMB}$ and $\Delta\rho_\text{CMB}$.   
In the upcoming section, we study the effects of modifying these two parameters on the neutrino oscillation probabilities for the following two cases:
\begin{itemize}
	\item 1-dimensional (1D) modification: $R_\text{CMB}$ is fixed at 3480 km and only $\Delta\rho_\text{CMB}$ is varied
	\item 2-dimensional (2D) modification: $R_\text{CMB}$ is varied simultaneously with $\Delta\rho_\text{CMB}$.
\end{itemize}
A 1D modification of $R_\text{CMB}$ is also possible, where $\Delta\rho_\text{CMB}$ is kept fixed. However, this modification has already been explored in Ref.~\cite{Upadhyay:2022jfd}, where the authors estimated the sensitivity to measure the location of the CMB using a three-layered profile. Additionally, the impact of changing the Earth's density model from a three layer to a five layer model on the measurement of $R_\text{CMB}$ is discussed in appendix \ref{app:sens_rcmb_measurement}.

We would like to highlight the fact that the density profile of Earth should have at least five layers if we want to explore the correlated effects of $R_\text{CMB}$ and $\Delta\rho_\text{CMB}$ while respecting the above-mentioned eight constraints. At the same time, a five-layered profile is sufficient to capture all the relevant features of the three-flavor neutrino oscillation probabilities in the presence of matter. The profile with higher number of layers does not change neutrino oscillation probabilities significantly. Therefore, in the present work, we use the five-layered density profile of Earth.

\begin{figure}[htp!]
	\centering
	\includegraphics[width=1.0\linewidth]{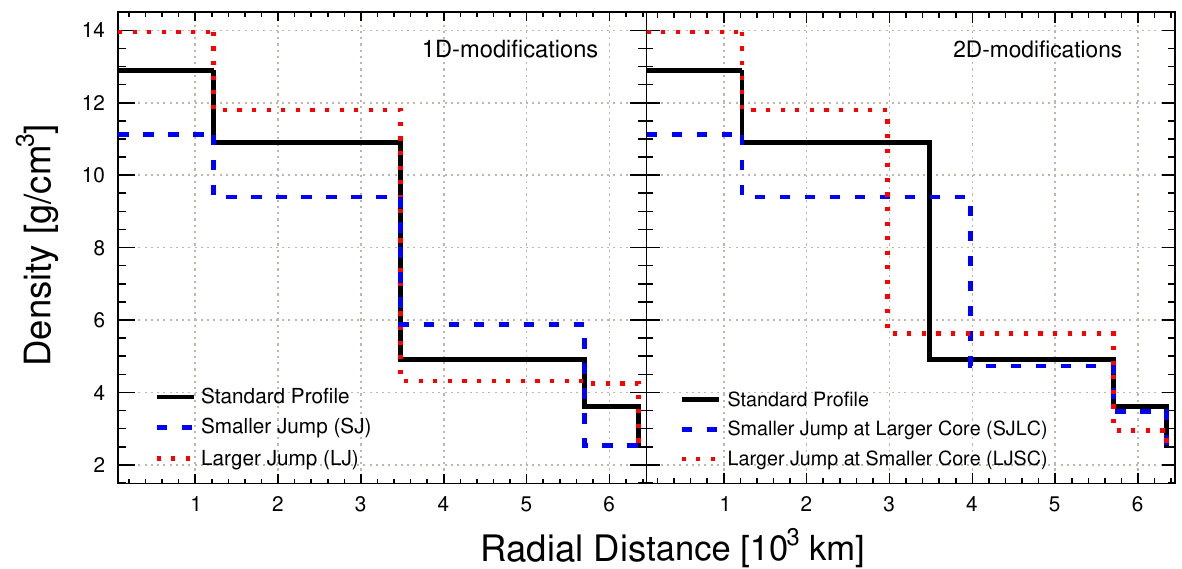}
	\mycaption{The five-layered density profiles of Earth as functions of the radial distances for some representative choices of density jumps at the location of the CMB, as mentioned in Table~\ref{tab:density_jump_variation}. The black curves correspond to the standard density profile, with a density jump of $\Delta \rho_\text{CMB} = $ 5.9965 g/cm$^3$ at the CMB with its standard location of $R_\text{CMB}$ = 3480 km. The left panel presents the 1D-modifications, where the dashed-blue (dotted-red) curve represents the SJ (LJ) scenario. The right panel corresponds to the 2D-modifications, where the dashed-blue (dotted-red) curve denotes the SJLC (LJSC) scenario.}
	\label{fig:dend_profile}
\end{figure} 

Table~\ref{tab:density_jump_variation} shows some benchmark values for the 1D-modifications, which correspond to the following two scenarios:
\begin{itemize}
	\item SJ: a smaller jump of $\Delta\rho_\text{CMB} = 3.5215$ g/cm$^3$,
	\item LJ: a larger jump of $\Delta\rho_\text{CMB} = 7.4807$ g/cm$^3$,
\end{itemize}
at the standard CMB, where $R_\text{CMB}$ is fixed at 3480 km. The density distributions for profiles with the standard jump, SJ, and LJ are shown in the left panel of Fig.~\ref{fig:dend_profile}.

The benchmark values for the 2D-modifications in Table~\ref{tab:density_jump_variation} correspond to the following two scenarios:
\begin{itemize}
	\item SJLC: a smaller jump of $\Delta\rho_\text{CMB} = 4.6662$ g/cm$^3$ at a larger core, with $\Delta R_\text{CMB} = + \, 500$ km, 
	\item LJSC: a larger jump of $\Delta\rho_\text{CMB} = 6.1691$ g/cm$^3$ at a smaller core, with $\Delta R_\text{CMB} = - \, 500$ km.
\end{itemize} 
The right panel of Fig.~\ref{fig:dend_profile} shows the density distributions for profiles with the standard jump at the standard CMB, SJLC, and LJSC. Now, we discuss the effects of modifying the density jump at the CMB and the location of the CMB on neutrino oscillation probabilities in the next section.

\subsection{Effects of modified density jumps and the location of CMB on oscillation probabilities}
\label{sec:density_jump_oscillation}

Atmospheric neutrinos and antineutrinos are produced about 15 km above the surface of Earth. They possess a wide range of energies from sub-GeV to a few TeV and travel over baselines ranging from about 15 km to 12757 km. While passing through Earth, the upward-going multi-GeV neutrinos encounter the Earth's matter effects due to the charged-current (CC) interactions with the ambient electrons. The effective matter potential is given by
\begin{equation}
V_\text{CC} = \pm\, \sqrt{2} G_F N_e \approx \pm \, 7.6 \times Y_e \times 10^{-14} \left[\frac{\rho}{\text{g/cm}^3}\right]~\text{eV}\,,
\end{equation}
where $N_e$ is the ambient electron number density, $G_F$ is the Fermi coupling constant, and $Y_e = N_e/(N_p + N_n)$ is the electron fraction of the matter having mass density $\rho$. Here, $N_p$ and $N_n$ denote the number densities of protons and neutrons inside matter. In our analysis, we consider $Y_e$ = 0.466 for the IC and OC (corresponding to a pure FeNi core) and $Y_e$ = 0.494 for the IM, MM, and OM (corresponding to a pyrolite mantle) as proposed in Ref.~\cite{Rott:2015kwa}. The $\pm$ signs correspond to neutrinos and antineutrinos, respectively. Due to these opposite signs for $V_\text{CC}$, oscillation probabilities for neutrinos and antineutrinos are modified differently inside matter. Thanks to the charge identification capability of the ICAL detector, it would be able to observe these different matter effects in neutrinos and antineutrinos separately, which is expected to enhance the sensitivity of ICAL for physics driven by matter effects. 

\begin{table}[htp!]
	\centering
	\begin{tabular}{|c|c|c|c|c|c|c|}
		\hline
		$\sin^2 2\theta_{12}$ & $\sin^2\theta_{23}$ & $\sin^2 2\theta_{13}$ & $\Delta m^2_\text{eff}$ (eV$^2$) & $\Delta m^2_{21}$ (eV$^2$) & $\delta_{\rm CP}$ & Mass Ordering\\
		\hline
		0.855 & 0.5 & 0.0875 & $2.49\times 10^{-3}$ & $7.4\times10^{-5}$ & 0 & NO\\
		\hline 
	\end{tabular}
	\mycaption{The benchmark values of neutrino oscillation parameters considered in this work. These values are consistent with the present global fits~\cite{Capozzi:2025wyn,NuFIT6.0,Esteban:2024eli,deSalas:2020pgw} to the neutrino data.}
	\label{tab:osc-param-value}
\end{table}

In this work, we perform the analysis using the benchmark values of the oscillation parameters given in Table~\ref{tab:osc-param-value}. Since we evaluate the sensitivities using 20 years of exposure, at this time scale, these parameters, along with the mass ordering, are expected to be measured precisely~\cite{Song:2020nfh}. Therefore, the values of these parameters are kept fixed, and the normal mass ordering (NO) is considered during the analysis, unless stated otherwise. The values of $\Delta m^2_{31}$ are obtained from the effective atmospheric mass-squared difference $\Delta m^2_\text{eff}$ using~\cite{deGouvea:2005hk,Nunokawa:2005nx}  
\begin{equation}
	\Delta m^2_\text{eff} = \Delta m^2_{31} - \Delta m^2_{21} (\cos^2\theta_{12} - \cos \delta_\text{CP} \sin\theta_{13}\sin2\theta_{12}\tan\theta_{23}).
	\label{eq:m_eff}
\end{equation}
For inverted mass ordering (IO), the sign of $\Delta m^2_\text{eff}$ is reversed, keeping its magnitude and the values of all other oscillation parameters unchanged.

The matter potential modifies the effective masses and flavor mixing of neutrinos~\cite{Wolfenstein:1977ue,Mikheev:1986gs,Mikheev:1986wj}. As a result, it alters the oscillation probabilities of neutrinos passing through Earth. The resonant enhancement of neutrino oscillation probabilities due to matter effects is known as the Mikheyev-Smirnov-Wolfenstein (MSW) resonance~\cite{Wolfenstein:1977ue,Mikheev:1986gs,Mikheev:1986wj}. The MSW resonance is significant in neutrinos for normal mass ordering, $m_1 < m_2 < m_3$, and in antineutrinos for inverted mass ordering, $m_3 < m_1 < m_2$. The MSW resonance depends upon the neutrino energy and the ambient electron density. In particular, neutrinos passing through the mantle feel the MSW resonance at energies of about $6-10$ GeV. In addition, the core-passing neutrinos propagate through two sharp jumps in density between the mantle and core at the CMB, which leads to significant changes in the oscillation probabilities at energies of about $3-6$ GeV. This phenomenon is known as the parametric resonance (PR)~\cite{Ermilova:1986,Akhmedov:1988kd,Krastev:1989,Akhmedov:1998ui,Akhmedov:1998xq} or neutrino oscillation length resonance (NOLR)~\cite{Petcov:1998su,Chizhov:1998ug,Petcov:1998sg,Chizhov:1999az,Chizhov:1999he}. The PR/NOLR also depends upon the density jump at the CMB and its location. We start by studying the effect of modification in the density jump at the CMB and its location on the neutrino oscillation probabilities.

\begin{figure}[htp!]
	\centering
	\includegraphics[width=0.85\linewidth]{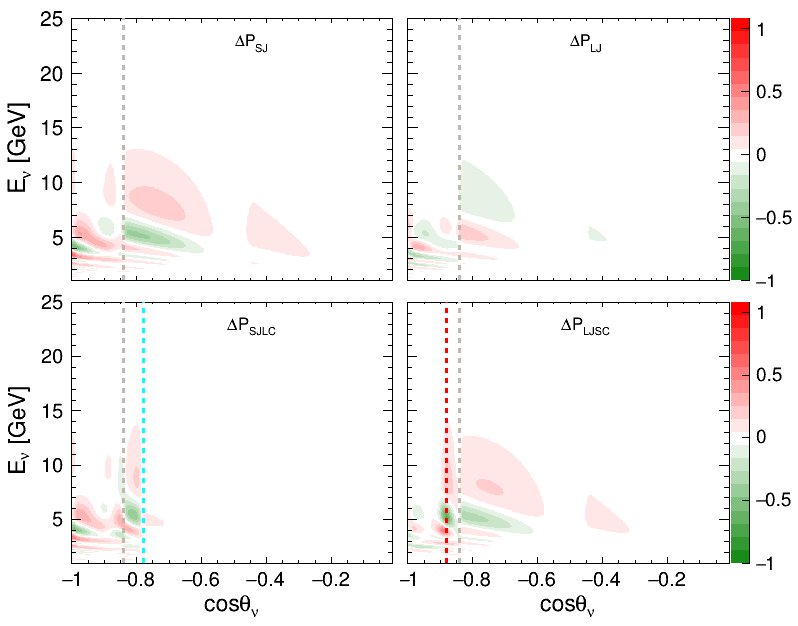}
	\mycaption{Oscillograms for the three-flavor $\nu_\mu$ survival probability difference, $\Delta P(\nu_\mu \rightarrow \nu_\mu)$, for the 1D (top panels) and 2D (bottom panels) modifications. The top left (right) panel corresponds to $\Delta P_\text{SJ}$ ($\Delta P_\text{LJ}$), as defined in Eq.~\ref{eq:deltaP_SJ} (Eq.~\ref{eq:deltaP_LJ}). The bottom left (right) panel corresponds to $\Delta P_\text{SJLC}$ ($\Delta P_\text{LJSC}$), as defined in Eq.~\ref{eq:deltaP_SJLC} (Eq.~\ref{eq:deltaP_LJSC}). The gray, red, and cyan vertical dotted lines represent the standard, smaller, and larger $R_\text{CMB}$ values, respectively.}
	\label{fig:prob_oscillogram_diff}
\end{figure} 

ICAL would be mainly sensitive to muon-type neutrino events, which have contributions from both the $\nu_\mu\rightarrow\nu_\mu$ disappearance channel and $\nu_e\rightarrow\nu_\mu$ appearance channel. However, more than 98\% of the contribution to $\nu_\mu$ would come from the disappearance channel. Therefore, we present the effects of modifications in density jump at the CMB and in its location relative to the standard values only for the $\nu_\mu$ survival probability, $P(\nu_\mu\rightarrow\nu_\mu)$, in the $(E_\nu, \cos\theta_\nu)$ plane. Neutrino oscillation patterns depend on both the neutrino energy and baseline. Figure~\ref{fig:prob_oscillogram_diff} presents the probability difference distributions between the 1D (2D) modifications and the standard scenario in the top (bottom) panels. The top left (right) panel shows the distribution of $\Delta P_\text{SJ}$ ($\Delta P_\text{LJ}$), and the bottom left (right) panel shows the distribution of $\Delta P_\text{SJLC}$ ($\Delta P_\text{LJSC}$), which are defined as 

\noindent
\begin{align}
	\Delta P_{\text{SJ}} &= P(\nu_{\mu} \rightarrow \nu_\mu)_{\text{SJ}} - P(\nu_{\mu} \rightarrow \nu_\mu)_{\text{standard}}\,,\label{eq:deltaP_SJ} \\
	\Delta P_{\text{LJ}} &= P(\nu_{\mu} \rightarrow \nu_\mu)_{\text{LJ}} - P(\nu_{\mu} \rightarrow \nu_\mu)_{\text{standard}}\,, \label{eq:deltaP_LJ}\\
	\Delta P_{\text{SJLC}} &= P(\nu_{\mu} \rightarrow \nu_\mu)_{\text{SJLC}} - P(\nu_{\mu} \rightarrow \nu_\mu)_{\text{standard}}\,,  \label{eq:deltaP_SJLC}\\
	\Delta P_{\text{LJSC}} &= P(\nu_{\mu} \rightarrow \nu_\mu)_{\text{LJSC}} - P(\nu_{\mu} \rightarrow \nu_\mu)_{\text{standard}}\,\label{eq:deltaP_LJSC}\,.
\end{align} 

The top panels of Fig.~\ref{fig:prob_oscillogram_diff} show that the probability differences in the 1D scenarios, SJ and LJ, are significant for core-passing neutrinos as well as for a fraction of the neutrinos that pass only through the mantle. This can be explained by the fact that the densities of both the core and the mantle are altered in the 1D-modifications (see Table~\ref{tab:density_jump_variation}). On the other hand, the bottom panels of Fig.~\ref{fig:prob_oscillogram_diff} for the 2D-modifications, SJLC and LJSC, show that the probability differences are significant mainly for neutrinos that pass through the core or very close to it.

From Fig.~\ref{fig:prob_oscillogram_diff}, we infer that the expected signal region would occur mainly at lower energies and larger baselines, corresponding to atmospheric neutrinos having trajectories with $\cos\theta_\nu \lesssim - \, 0.8$. A detector like ICAL, with the help of its excellent angular resolution in the multi-GeV range of energy, would be able to probe the density jump and the CMB radius simultaneously.

\section{Atmospheric neutrinos at INO-ICAL}
\label{sec:event_genration}

The proposed 50 kt iron calorimeter (ICAL) detector at the India-based Neutrino Observatory (INO)~\cite{ICAL:2015stm} is designed to detect multi-GeV atmospheric neutrinos and antineutrinos separately over a wide range of baselines. ICAL would have three modules of size 16 m $\times$ 16 m $\times$ 14.5 m, each having about 151 layers of iron plates of thickness 5.4 cm. The iron layers would be vertically stacked with a gap of about 4 cm, where the glass Resistive Plate Chambers (RPCs) of size 2 m $\times$ 2 m would be inserted. The iron layers would act as a passive detector element, providing a target material for the neutrino interactions, whereas the RPCs would act as an active detector element which would detect the secondary charged particles produced in neutrino interactions.

The charged-current interactions of muon neutrinos ($\nu_\mu$) and antineutrinos ($\bar{\nu}_\mu$) produce $\mu^-$ and $\mu^+$, respectively. Muons are minimum ionizing particles in the multi-GeV energy range, and hence, they deposit energy in the detector in the form of a long track. The magnetic field of about 1.5 T~\cite{Behera:2014zca} would enable ICAL to distinguish $\mu^-$ and $\mu^+$ by observing the direction of curvature of the muon track, and hence, to identify the parent $\nu_\mu$ or $\bar{\nu}_\mu$. This charge identification capability of ICAL would play a crucial role in analyses driven by matter effects, such as the determination of the neutrino mass ordering, the measurement of the octant of $\theta_{23}$, and neutrino oscillation tomography. The time resolution of RPCs at the ns level~\cite{Dash:2014ifa,Bhatt:2016rek,Gaur:2017uaf} would help ICAL to distinguish the upward-going and downward-going events. In the energy range of $1-25$ GeV, ICAL would be able to measure the direction of a muon $(\cos\theta_\mu)$ with an excellent angular resolution of about 1$^\circ$~\cite{Chatterjee:2014vta} and the energy of a muon ($E_\mu$) with a resolution of about $10-15$\%~\cite{Chatterjee:2014vta}.

In the multi-GeV range of energies, neutrinos can also undergo deep-inelastic scattering (DIS), which produce hadrons along with leptons. The hadrons deposit energy in the detector in the form of a shower. The hadrons can carry a significant fraction of the incoming neutrino energy ($E_\nu$). The hadron energy deposited in the detector can be defined as ${E'}_\text{had} = E_\nu - E_\mu$. ICAL would be able to measure the hadron energy with a resolution of about 40\% for ${E'}_\text{had} > 5$ GeV~\cite{Devi:2013wxa}.

In the present work, the unoscillated atmospheric neutrino events at ICAL are simulated using the NUANCE~\cite{Casper:2002sd} Monte Carlo (MC) neutrino event generator, which takes as inputs the ICAL geometry and the Honda 3D flux of atmospheric neutrinos at the proposed INO site at Theni~\cite{Athar:2012it,Honda:2015fha}. The effect of solar modulation on the atmospheric neutrino flux has been taken into account by using a neutrino flux with high solar activity for half of the exposure and with low solar activity for the other half. A rock overburden of at least 1 km (3800 m water equivalent) from all directions at the INO site would reduce the cosmic muon background by a factor of about $10^6$~\cite{Dash:2015blu}. Additionally, to veto muon tracks entering from outside the detector, analyses at ICAL consider only those events for which the event vertices lie completely inside the detector and far from the edges. Therefore, at ICAL, we expect negligible background due to downward-going cosmic muons. The interactions of tau neutrinos produce tau leptons, which can decay into muons. The number of these muon events is only about 2\% of the total number of upward-going muons from $\nu_{\mu}$ interactions, and most of them have energies below the energy threshold (1 GeV) of ICAL. Therefore, we do not consider muon events from tau decays in the present analysis. Since we evaluate median sensitivities in this analysis, we generate the unoscillated MC neutrino events for a large exposure of about 1000 years to suppress the statistical fluctuations.  We incorporate the three-flavor neutrino oscillations in the presence of Earth's matter effects using a reweighting algorithm~\cite{Devi:2014yaa,Ghosh:2012px,Thakore:2013xqa}.

\begin{figure}[htp!]
	\centering
	\includegraphics[width=\linewidth]{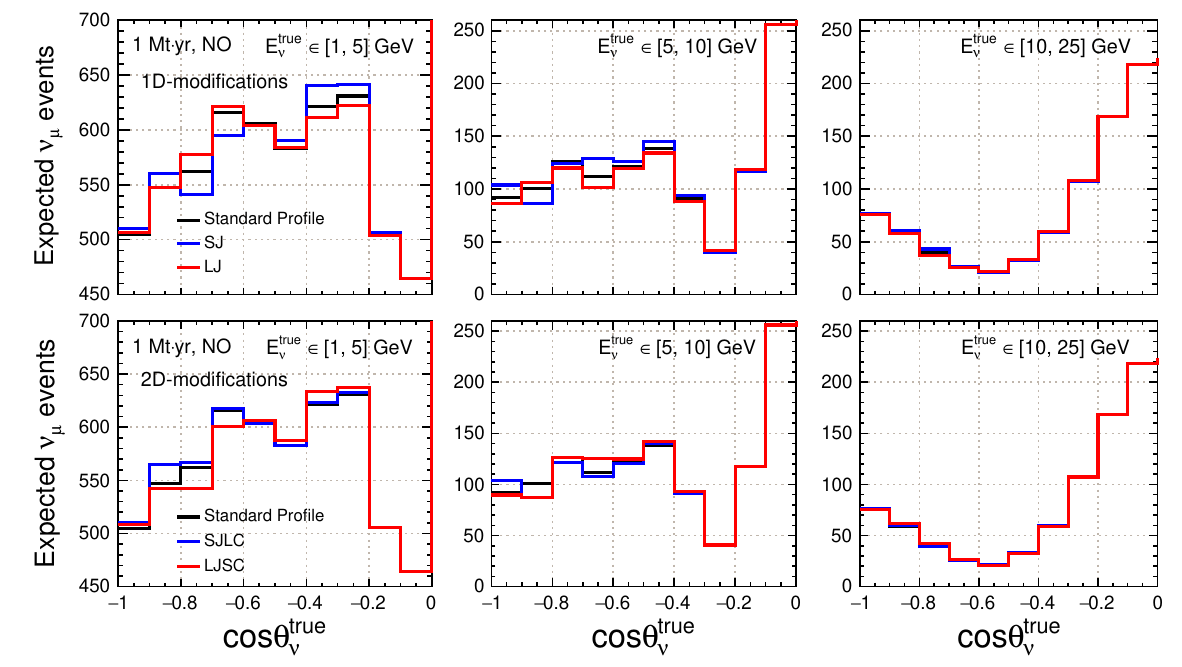}
	\mycaption{The expected upward-going muon neutrino event distributions as a function of the true neutrino arrival direction ($\cos\theta^\text{true}_{\nu}$) at the ICAL detector. The top (bottom) row corresponds to the 1D (2D) modification scenarios. In each row, the black curves represent the standard case, while the blue and red curves in the top (bottom) panels correspond to the SJ and LJ (SJLC and LJSC) scenarios, respectively, for different energy ranges. These neutrino events are obtained before applying detector reconstruction efficiencies and resolution effects.}
	\label{fig:true_events}
\end{figure} 

Figure~\ref{fig:true_events} shows the distribution of upward-going muon neutrino events as a function of the true neutrino arrival direction ($\cos\theta^\text{true}_{\nu}$), expected at the ICAL detector for an exposure of 1 Mt$\cdot$yr, assuming NO. We compare event distributions for the 1D (2D) modifications with the standard case in the top (bottom) row. These neutrino events are obtained before applying detector reconstruction efficiencies and resolution effects. The impacts of both the 1D and 2D modifications on neutrino events relative to the standard case are most evident for neutrino energies $E^\text{true}_\nu < 10$ GeV and for longer baselines ($\cos\theta^\text{true}_\nu < - \, 0.4$). This corresponds to the region where Earth's matter effects on neutrino oscillations are prominent, as also visible in Fig.~\ref{fig:prob_oscillogram_diff}.

The detector responses for muons~\cite{Chatterjee:2014vta} and hadrons~\cite{Devi:2013wxa} have been incorporated using the migration matrices provided by the ICAL collaboration. The reconstruction efficiency, CID efficiency, energy resolution, and angular resolution  of the ICAL detector for reconstructed muons are mentioned in figures 13, 14, 11, and 6, respectively, of Ref.~\cite{Chatterjee:2014vta}. After incorporating the detector responses, we obtain reconstructed observables such as the muon energy ($E_\mu^\text{rec}$), the muon direction ($\cos\theta_\mu^\text{rec}$), and the hadron energy (${E^\prime}_{\text{had}}^\text{rec}$)~\cite{Devi:2014yaa}.

\begin{table}[htp!]
	\centering
	\begin{tabular}{| c | c | c | c | c | c |} 
		\hline \hline
		&  \multicolumn{2}{c|}{1D modifications} & 
		\multicolumn{2}{c|}{2D modifications} & Standard\\
		\cline{2-5}
		&  SJ & LJ & SJLC & LJSC & Case\\ 
		\hline 
		$\mu^-$ events & 8865 & 8835 & 8863 & 8840 & 8848 \\
		\hline
		$\mu^+$ events & 4030 & 4038 & 4034 & 4032 & 4033 \\ 
		\hline \hline
	\end{tabular}
	\mycaption{The total expected number of reconstructed $\mu^-$ and $\mu^+$ events using 1 Mt$\cdot$yr of exposure of the ICAL detector for the 1D-modifications, 2D-modifications, and the standard case.}
	\label{tab:events}
\end{table}

For the evaluation of sensitivities, the reconstructed events are scaled down from the 1000-year MC to a 20-year MC, which corresponds to 1 Mt$\cdot$yr of exposure of ICAL. In 20 years, ICAL is expected to observe about 8848 (4033) reconstructed $\mu^-$ ($\mu^+$) events, considering the three-flavor neutrino oscillations with Earth's matter effects and assuming the standard five-layered profile of Earth for NO. In Table~\ref{tab:events}, we show the expected number of reconstructed $\mu^-$ and $\mu^+$ events for 1 Mt$\cdot$yr of exposure of the ICAL detector for all four scenarios of the 1D and 2D modifications, viz. SJ, LJ, SJLC, and LJSC. It may be observed that the total number of events for these four scenarios are not significantly different. However, the good directional and energy resolution of the ICAL detector permits us to bin the expected events in terms of the reconstructed observables $E_\mu^\text{rec}$, $\cos\theta_\mu^\text{rec}$, and ${E^\prime}_{\text{had}}^\text{rec}$. The binned event distributions for these scenarios are expected to be significantly different and would give rise to the sensitivity of ICAL to distinguish among them.

\begin{figure}[htp!]
	\centering
	\includegraphics[width=0.85\linewidth]{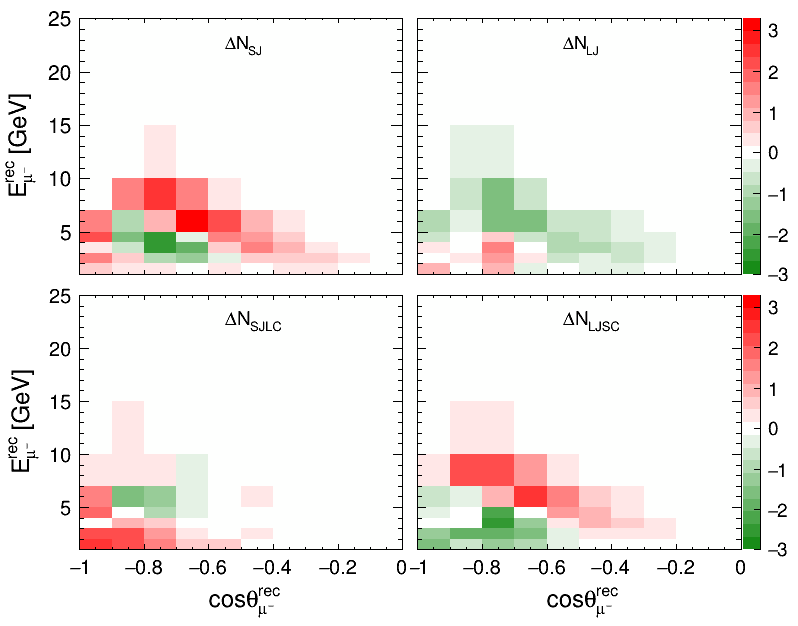}
	\mycaption{Distributions of reconstructed $\mu^-$ event differences between the modified density jump and the standard density jump in the plane of ($E^\text{rec}_{\mu^-}$, $\cos\theta^\text{rec}_{\mu^-}$) for 1 Mt$\cdot$yr of exposure at ICAL. In the top left (right) panel $\Delta N_\text{SJ}$ ($\Delta N_\text{LJ}$) represents the $\mu^-$ event differences between the SJ (LJ) scenario and the standard density jump. In the bottom left (right) panel, $\Delta N_\text{SJLC}$ ($\Delta N_\text{LJSC}$) denotes the $\mu^-$ event differences between the SJLC (LJSC) scenario and the standard density jump.}
	\label{fig:event_diff}
\end{figure} 

Now, we explore the impact of density jump variations at the standard $R_\text{CMB}$ (1D-modifications) and at the modified $R_\text{CMB}$ (2D-modifications) on the bin-wise distribution of reconstructed muon events at the ICAL detector. In order to obtain the distributions of event differences in Fig.~\ref{fig:event_diff}, we use a binning scheme with 9 bins in $E^\text{rec}_{\mu^-}$ consisting of 4 bins of 1 GeV in ($1-5$) GeV, 1 bin of 2 GeV in ($5-7$) GeV, 1 bin of 3 GeV in ($7-10$) GeV, and 3 bins of 5 GeV in ($10-25$) GeV, and 20 uniform bins for $\cos\theta^\text{rec}_{\mu^-}$ in the range of $-\,1$ to 1. The events have been integrated over the observable ${E^\prime}_{\text{had}}^\text{rec}$ in the range of 0 to 25 GeV.  Since for normal mass ordering, Earth's matter effects are significant for neutrinos, we choose to present the distributions of event differences only for reconstructed $\mu^-$ events. The reason for using this coarser binning scheme in Fig.~\ref{fig:event_diff} is to visually highlight the expected signal region in the ($E^\text{rec}_{\mu^-}$, $\cos\theta^\text{rec}_{\mu^-}$) plane. The coarser bins show clear boundaries of the prominent regions which are easy to recognise by eye, something that is not easily achieved with the finer analysis binning discussed in Section~\ref{sec:binning}.

The top (bottom) panels of Fig.~\ref{fig:event_diff} show the distribution of event differences between the modified density jumps for the 1D (2D) modifications and the standard density jump, in the plane of ($E^\text{rec}_{\mu^-}$, $\cos\theta^\text{rec}_{\mu^-}$) for 1 Mt$\cdot$yr of exposure. The top left (right) panel presents the distribution of event differences between the SJ (LJ) scenario and the standard density jump, while the bottom left (right) panel shows the distribution of event differences between the SJLC (LJSC) scenario and the standard density jump. 

For the 1D-modifications, the event differences are observed to be significant for the core-passing neutrinos as well as for low-energy neutrinos that cross only the mantle, whereas the event differences are more prominent only for the core-passing neutrinos for the 2D modifications. The smearing of these differences over a wider region in Fig.~\ref{fig:event_diff} as compared to Fig.~\ref{fig:prob_oscillogram_diff} is mostly the result of the angular smearing caused by the deviation of the direction of the reconstructed muon with respect to the true direction of the incoming neutrino.

\section{Analysis method}
\label{sec:analysis_metho}

\subsection{Binning scheme}
\label{sec:binning}

\begin{table}[htp!]
	\centering
	\begin{tabular}{|c|c|c|c c|}
		\hline 
		\hline
		Observable & Range & Bin width & \multicolumn{2}{c|}{Number of bins} \\ 
		\hline 
		\multirow{4}{*}{$E_\mu^{\rm rec}$ (GeV)}  & [1, 6] & 0.5 & 10& 
		\rdelim\}{4}{7mm}[16] \cr
		& [6, 12] & 2 & 3 &\cr
		& [12, 15] & 3 & 1 & \cr
		& [15, 25] & 5 & 2 & \cr
		\hline
		\multirow{4}{*}{$\cos\theta^{\text{rec}}_\mu$} & [-1.0, -0.85] & 0.0125 & 12 & \rdelim\}{4}{7mm}[39]\cr
		& [-0.85, -0.4] & 0.025 & 18 &\cr
		& [-0.4, 0] & 0.1 & 4 & \cr
		& [0, 1] & 0.2 & 5 & \cr
		\hline
		\multirow{3}{*}{$E^{\prime \text{rec}}_{\text{had}}$ (GeV)}  & [0, 2] & 1 & 2  & \rdelim\}{3}{7mm}[4] \cr
		& [2, 4] & 2 & 1 &\cr
		& [4, 25] & 21 & 1 &\cr
		\hline 
		\hline
		
	\end{tabular}
	\mycaption{The binning scheme adopted for our analyses, which is the same for both $\mu^-$ and $\mu^+$ events.}
	\label{tab:binning_scheme}
\end{table}

In this section, we present the binning scheme considering the observables $E_\mu^{\rm rec}$, $\cos\theta^{\text{rec}}_\mu$, and $E^{\prime \text{rec}}_{\text{had}}$. This binning scheme, as used in Ref.~\cite{Upadhyay:2022jfd} and shown in Table~\ref{tab:binning_scheme}, includes a total of 16 bins for $E_\mu^{\rm rec}$ considered in the range of $1 - 25$ GeV, 39 bins for $\cos\theta^{\text{rec}}_\mu$ from $-\,1$ to 1, and 4 bins for $E^{\prime \text{rec}}_{\text{had}}$ spanning the range of $0-25$ GeV. It ensures a sufficient number of events in each bin. We use the same binning scheme across different scenarios with varying $R_\text{CMB}$ due to the limited neutrino sensitivity to such variations. We adopt the same binning scheme for both $\mu^-$ and $\mu^+$ events.

\subsection{Numerical analysis}
\label{sec:chi2}

In order to estimate the expected median sensitivity of ICAL to the correlated constraints on the density jumps and the location of the CMB, we perform a $\chi^2$ analysis within the frequentist approach~\cite{Blennow:2013oma}. We use a binning scheme for matter-effect studies as discussed in section~\ref{sec:binning}. The Poissonian $\chi^2_-$~\cite{Baker:1983tu} for $\mu^-$ is defined in terms of the reconstructed observables $E_\mu^\text{rec}$, $\cos\theta_\mu^\text{rec}$, and ${E^\prime}_{\text{had}}^\text{rec}$ as considered in Ref~\cite{Devi:2014yaa}:
\begin{equation}\label{eq:chisq_mu-}
\chi^2_- = \mathop{\text{min}}_{\xi_l} \left(\sum_{i=1}^{N_{{E'}_\text{had}^\text{rec}}} \sum_{j=1}^{N_{E_{\mu}^\text{rec}}} \sum_{k=1}^{N_{\cos\theta_\mu^\text{rec}}} \left[2(N_{ijk}^\text{theory} - N_{ijk}^\text{data}) -2 N_{ijk}^\text{data} \ln\left(\frac{N_{ijk}^\text{theory} }{N_{ijk}^\text{data}}\right)\right] + \sum_{l = 1}^5 \xi_l^2 \right) \,,
\end{equation}
with 
\begin{equation}
N_{ijk}^\text{theory} = N_{ijk}^0\left(1 + \sum_{l=1}^5 \pi^l_{ijk}\xi_l\right)\,.
\label{eq:chisq_2}
\end{equation}
Here, $N_{ijk}^\text{theory}$ and $N_{ijk}^\text{data}$ denote the number of expected and observed reconstructed $\mu^-$ events in a particular ($E^{\text{rec}}_\mu$, $\cos\theta^{\text{rec}}_\mu$, $E^{\prime \text{rec}}_{\text{had}}$) bin, respectively. The numbers $N_{E_{\mu}^\text{rec}}$, $N_{\cos\theta_\mu^\text{rec}}$, and $N_{{E'}_\text{had}^\text{rec}}$ denote the total number of bins for each of the respective observables. The variable $N_{ijk}^0$ represents the number of expected events in a given bin without considering systematic uncertainties. In this work, we incorporate the following five systematic uncertainties~\cite{Ghosh:2012px,Thakore:2013xqa} using the well-known method of pulls~\cite{Gonzalez-Garcia:2004pka,Huber:2002mx,Fogli:2002pt}: (i) 20\% flux normalization uncertainty, (ii) 5\% energy dependent tilt error in flux, (iii) 5\% zenith angle dependent tilt error in flux (iv) 10\% uncertainty on cross section, and (v) 5\% overall systematics. The pull variables for systematic uncertainties are given by $\xi_l$ in Eqs.~\ref{eq:chisq_mu-} and ~\ref{eq:chisq_2}.

The $\chi^2_+$ for reconstructed $\mu^+$ events is estimated following the same procedure as mentioned above for $\mu^-$ events. To estimate the resultant median sensitivity of the ICAL detector, we add the individual contributions from both $\chi^2_-$ and $\chi^2_+$ to get the total $\chi^2$:
\begin{equation}
\chi^2 = \chi^2_- + \chi^2_+\,.
\end{equation}
The MC data for this analysis is simulated using the benchmark values of oscillation parameters given in Table~\ref{tab:osc-param-value} as the true parameters. In the fit, we minimize the total $\chi^2$ over the pull variables $\xi_l$, while keeping the oscillation parameters fixed at their benchmark values.

\section{Results}
\label{sec:results}

In this section, we present the sensitivity of the proposed ICAL detector to the density jumps and the location of the CMB. We simulate the MC data assuming the five-layered density profile of Earth, with the standard density jumps and the standard radii of the IC, OC, IM, OM, and the Earth.

\subsection{Constraining correlated density jumps assuming the standard CMB location}

We quantify the expected sensitivity of the ICAL detector for measuring the density jumps while keeping the standard location of the CMB fixed (1D-modifications) as follows:
\begin{equation}
\Delta \chi^2_{\text{1D-DJ}} = \chi^2(\text{modified } \Delta\rho) - \chi^2 (\text{standard } \Delta\rho)\,,
\label{eq:delta_chisq_1D}
\end{equation}
where DJ stands for ``density jump'', which is denoted by $\Delta\rho$. The sensitivity evaluated using $\Delta \chi^2_{\text{1D-DJ}}$ corresponds to a correlated measurement of the four density jumps at the IC-OC, OC-IM (CMB), IM-OM, and OM-crust boundaries\footnote{As shown in Section~\ref{sec:density_jump_variation}, these four jumps are completely correlated in our modification procedure. This implies that, given any one of these density jumps, the other three are uniquely determined.}. Clearly, $\Delta\rho$ can be interpreted in terms of $\Delta\rho_\text{IC-OC}$, $\Delta\rho_\text{CMB}$, $\Delta\rho_\text{IM-OM}$, or $\Delta\rho_\text{OM-crust}$. 

\begin{figure}[htp!]
	\centering
	\includegraphics[width=0.495\linewidth]{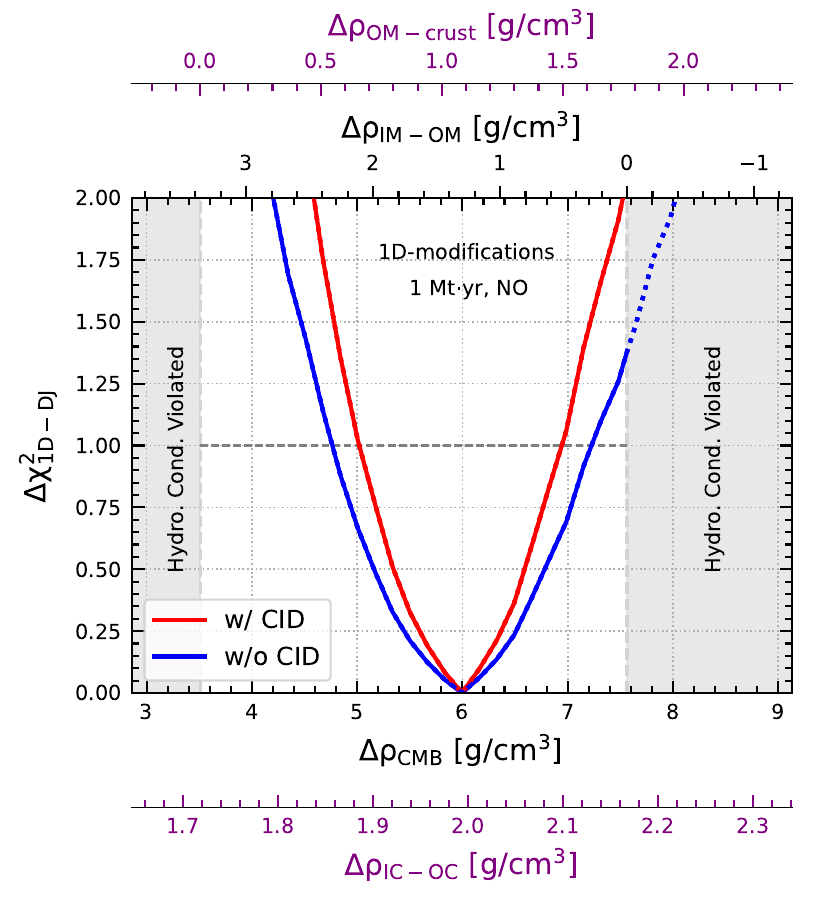}
	\includegraphics[width=0.495\linewidth]{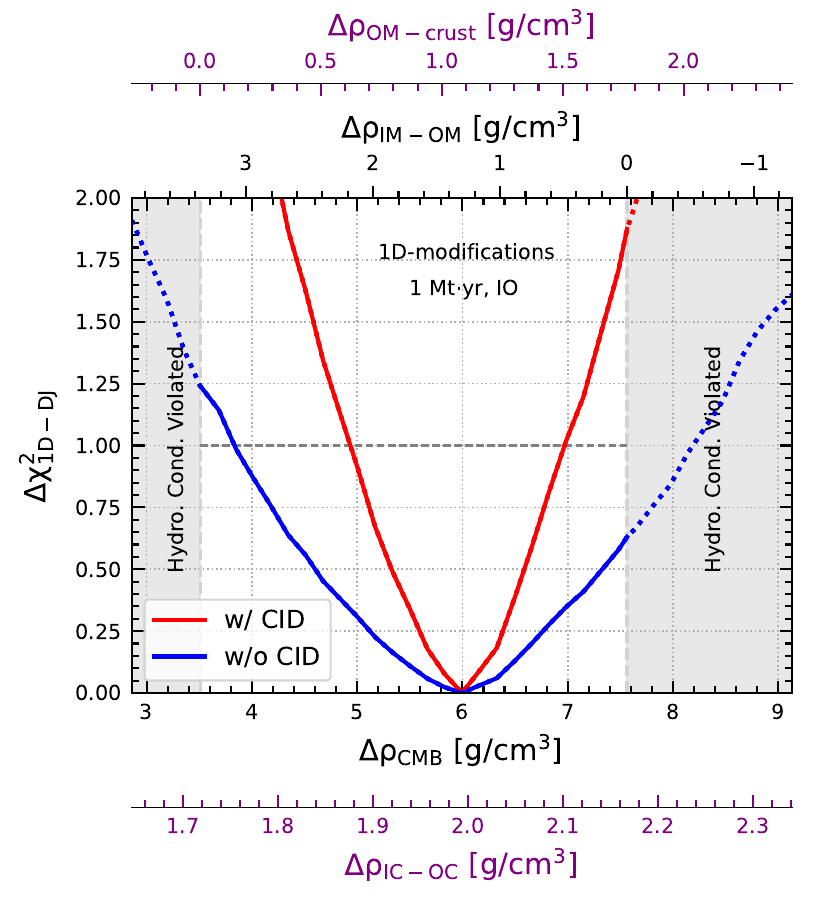}
	\mycaption{The median $\Delta\chi^2_\text{1D-DJ}$ as a function of the density jumps at the OM-crust and IM-OM (CMB and IC-OC) boundaries, corresponding to the top (bottom) x-axis, for the 1D-modifications. The gray area represents the unphysical region where the hydrostatic equilibrium condition is violated. The left (right) panel presents the sensitivity for NO (IO). The red (blue) curve corresponds to the sensitivity with (without) the CID capability of ICAL.}
	\label{fig:1D_results}
\end{figure}

Figure~\ref{fig:1D_results} shows the expected sensitivity of the ICAL detector, expressed in terms of $\Delta \chi^2_{\text{1D-DJ}}$, as a function of the density jumps at the corresponding boundaries with an exposure of 1 Mt$\cdot$yr. The left (right) panel of Fig.~\ref{fig:1D_results} presents the expected sensitivity assuming NO (IO). We present the sensitivity results in terms of the four x-axes corresponding to the above-mentioned four density jumps. The top two x-axes represent the density jumps at the OM-crust and the IM-OM boundaries, while the bottom two x-axes correspond to that at the CMB and the IC-OC boundaries. The gray areas represent the unphysical parameter space in which the hydrostatic equilibrium condition is violated. We can observe that some parts of the red and the blue curves lie in the gray regions, which are denoted by the dotted curves. From the red curve in the left panel of Fig.~\ref{fig:1D_results}, we can observe that the ICAL detector would be able to constrain the value of the density jump at the CMB ($\Delta\rho_\text{CMB}$) to the range [5.02, 6.95] g/cm$^3$ at $1\sigma$. This corresponds to a precision of about 15\%. The CID capability of the ICAL detector would play an important role in achieving this precision. In the absence of CID capability, the sensitivity range would deteriorate to [4.76, 7.23] g/cm$^3$ at $1\sigma$, which corresponds to a precision of about 20\% (for an insight into where this deterioration comes from, see the top panels of Fig.~\ref{fig:chi2_distribution} in appendix~\ref{app:sen_regions}).

From the red curve in the right panel of Fig.~\ref{fig:1D_results}, we can observe that, if the true mass ordering is inverted, ICAL would be able to constrain the value of $\Delta \rho_\text{CMB}$ to the range [4.94, 6.97] g/cm$^3$. This is close to the constraint obtainable in the case of true normal ordering without CID capability. However, without CID capability, the sensitivity is reduced significantly. We have checked that these sensitivities remain almost the same even if we minimize over the relevant oscillation parameters in their allowed ranges in the fit.

\subsection{Constraining correlated density jumps and location of CMB }

\begin{figure}[htp!]
	\centering
	\includegraphics[width=0.45\linewidth]{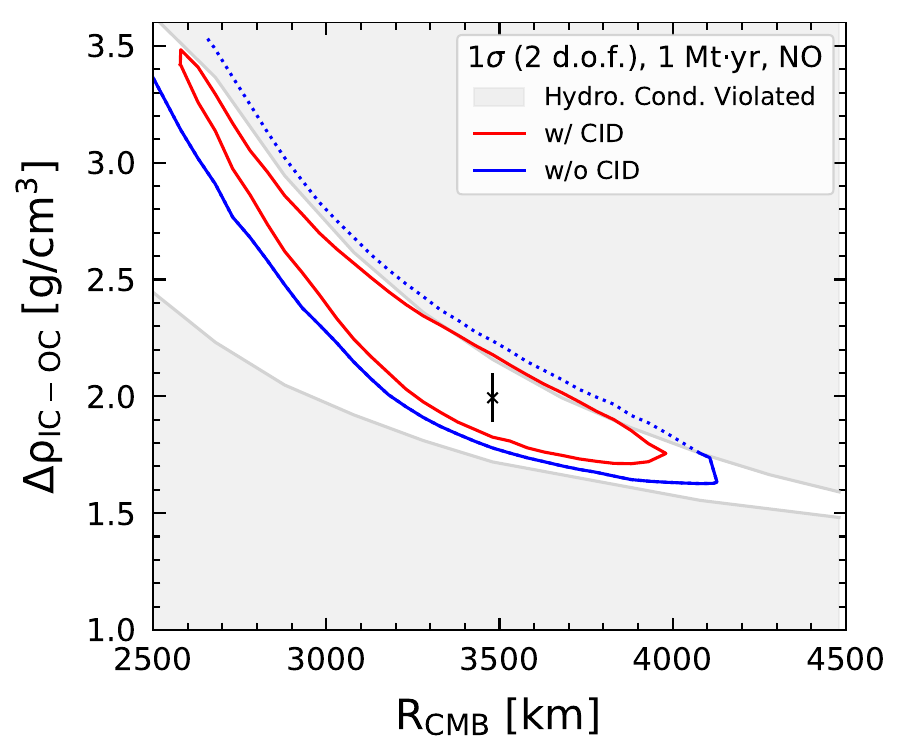}
	\includegraphics[width=0.45\linewidth]{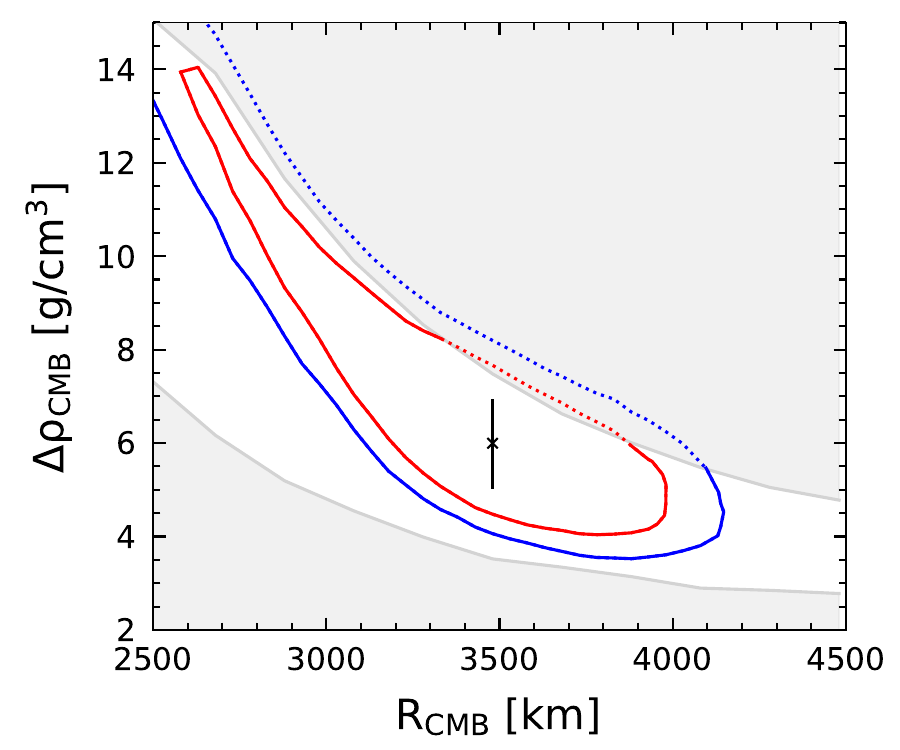}
	\includegraphics[width=0.45\linewidth]{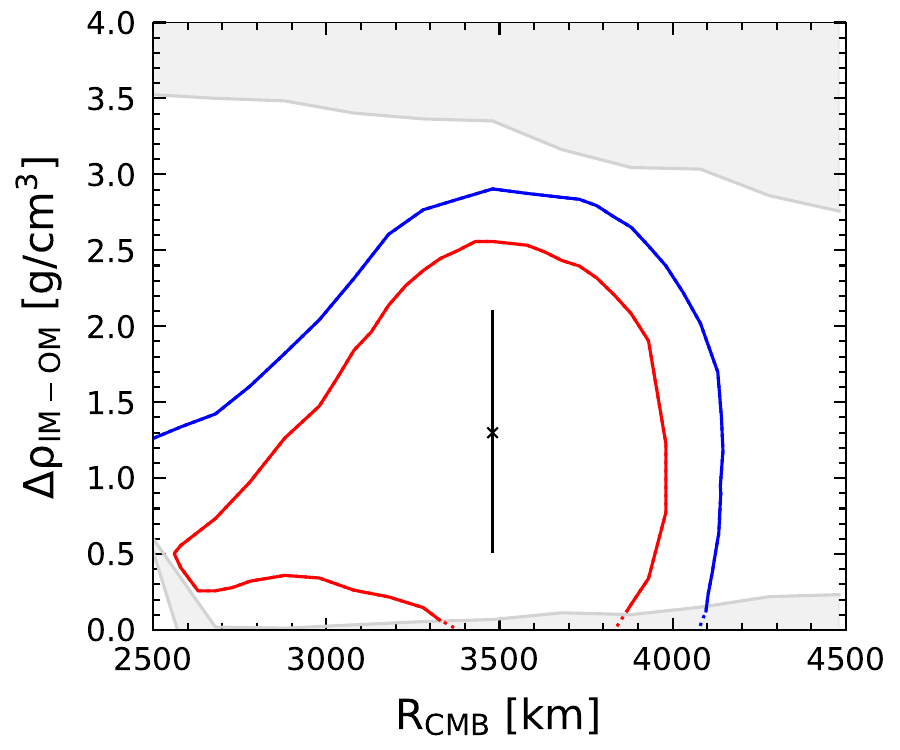}
	\includegraphics[width=0.45\linewidth]{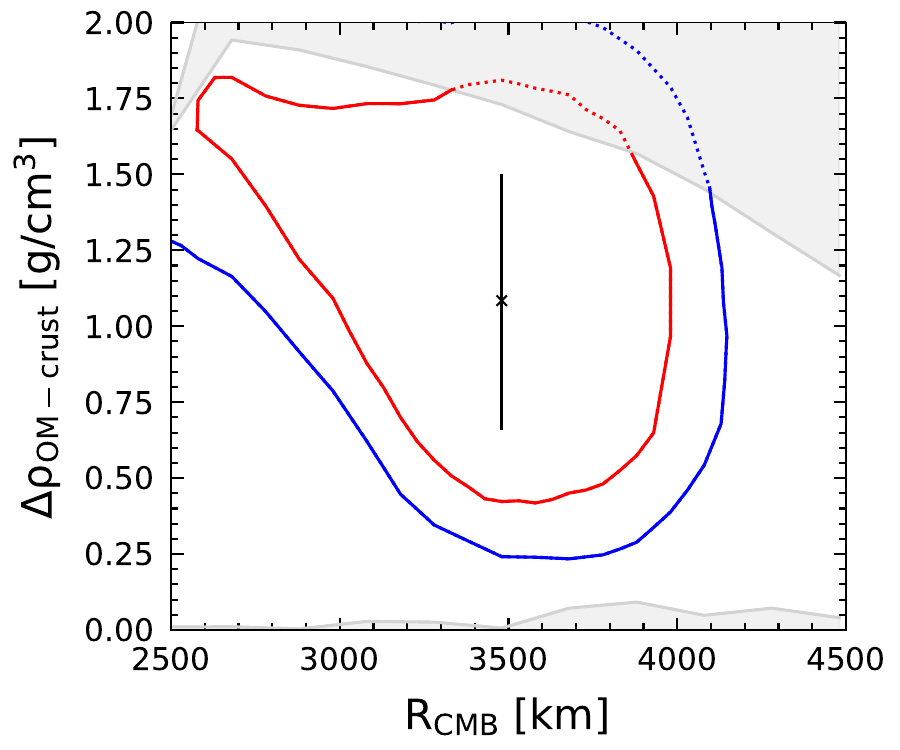}
	\mycaption{The median $\Delta\chi^2_\text{DJ-CMB}$ sensitivity contours in the plane of the density jumps and the location of the CMB radius at $1\sigma$ (2 d.o.f.) for 1 Mt$\cdot$yr of exposure. The red (blue) contour corresponds to the sensitivity with (without) the CID capability of the ICAL detector. The gray areas represent the unphysical region where the hydrostatic equilibrium condition is violated. The black crosses and vertical bars represent the standard values and $1\sigma$ constraints for the density jumps, respectively, assuming standard $R_\text{CMB} = 3480$ km for 1D modifications as shown in Fig.~\ref{fig:1D_results}.}
	\label{fig:2D_results}
\end{figure}

In this section, we present the expected sensitivities for the 2D-modifications, where the correlated density jumps and the location of the CMB radius are modified simultaneously. The ICAL sensitivity is quantified as
\begin{equation}
\Delta \chi^2_{\text{DJ-CMB}} = \chi^2(\text{modified } \{\Delta\rho, R_\text{CMB}\}) - \chi^2 (\text{standard } \{\Delta\rho, R_\text{CMB}\})\,.
\label{eq:delta_chisq_2D}
\end{equation}
The expected sensitivities of the ICAL detector for constraining the correlated density jumps and the location of the CMB radius simultaneously are shown in Fig.~\ref{fig:2D_results} in terms of contours at $1\sigma$ (2 d.o.f.) for the exposure of 1 Mt$\cdot$yr, assuming normal mass ordering. The top left (right) panel shows the sensitivity contours for simultaneously constraining the density jump at the IC-OC (CMB) boundary and the location of the CMB radius, whereas the bottom left (right) panel presents the sensitivity for the density jump at the IM-OM (OM-crust) boundary and the CMB radius.

For each plot, it is observed that some parts of the red and the blue contours are present in the gray region, which are denoted by the dotted curves. Note that, in the context of the five-layered model of Earth, the white region in Fig.~\ref{fig:2D_results} is allowed by the measurements of $M_\text{E}$, $I_\text{E}$, the density $\rho_\text{crust}$, the radii $R_\text{IC}$, $R_\text{IM}$, $R_\text{OM}$, $R_\text{E}$, and the density ratio $\rho_\text{IC}/\rho_\text{OC}$. In this study, we do not include any prior information on $\Delta\rho_\text{CMB}$ and $R_\text{CMB}$ because our objective is to use neutrino oscillation data alone to probe the correlated constraints on these two parameters. In certain regions of Fig.~\ref{fig:2D_results}, the constraints imposed by the hydrostatic equilibrium condition are observed to be stronger than the sensitivity of neutrino oscillation data itself. Incorporating the hydrostatic equilibrium condition as an additional constraint would further strengthen the net bounds obtained by the neutrino data in such regions. This clearly highlights the complementarity of our proposed method with existing measurements of the above quantities in the long run.

\subsection{Impact of different true choices of $\sin^2\theta_{23}$}
\label{sec:results_impact_true_theta23}

So far, we have considered $\sin^2\theta_{23}$ (true) = 0.5 in our analysis. However, the present global fits indicate that $\theta_{23}$ may not correspond to maximal mixing ($\sin^2\theta_{23}=0.5$). It can lie either in the lower octant $(\sin^2\theta_{23} < 0.5)$ or the higher octant ($\sin^2\theta_{23} > 0.5$). Therefore, in this section, we discuss how the ICAL sensitivity for constraining the density jump and the CMB radius may change if $\theta_{23}$ (true) is non-maximal.

\begin{figure}[htp!]
	\centering
	\includegraphics[width=0.65\linewidth]{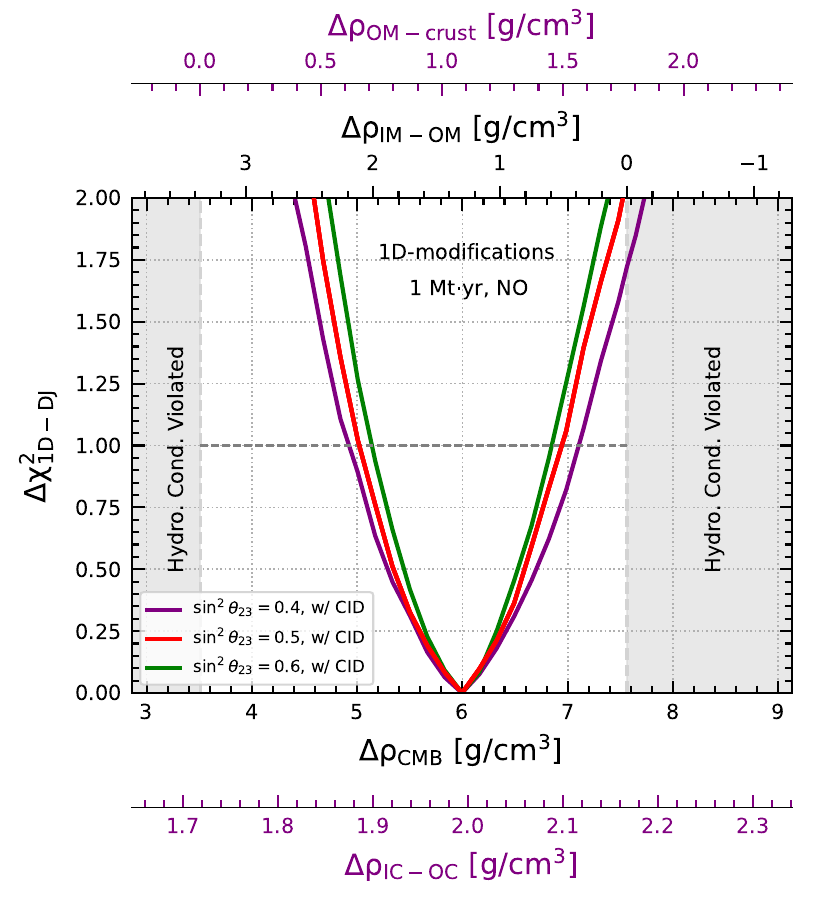}
	\mycaption{The median $\Delta\chi^2_\text{1D-DJ}$ as a function of the density jumps at the corresponding boundaries for the 1D modifications, with the CID capability of the ICAL detector. The sensitivities are evaluated for three choices of the true value of $\sin^2\theta_{23}$ = 0.4, 0.5, and 0.6, corresponding to the violet, red, and green curves, respectively. The other details are the same as in Fig.~\ref{fig:1D_results}.}
	\label{fig:1D_results_impact_theta23}
\end{figure}

\begin{figure}[htp!]
	\centering
	\includegraphics[width=0.45\linewidth]{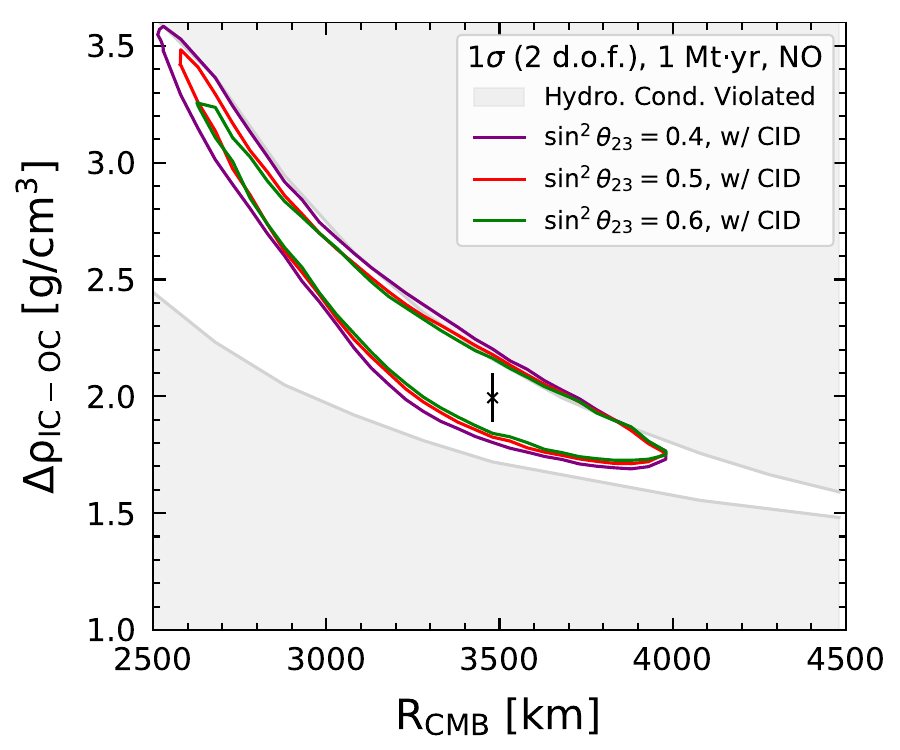}
	\includegraphics[width=0.45\linewidth]{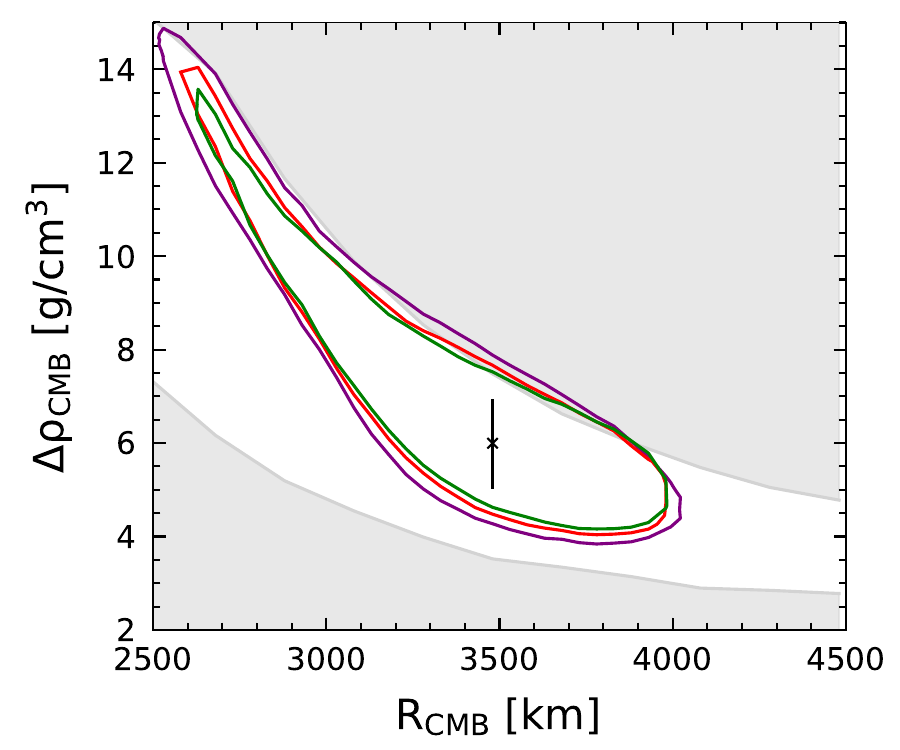}
	\includegraphics[width=0.45\linewidth]{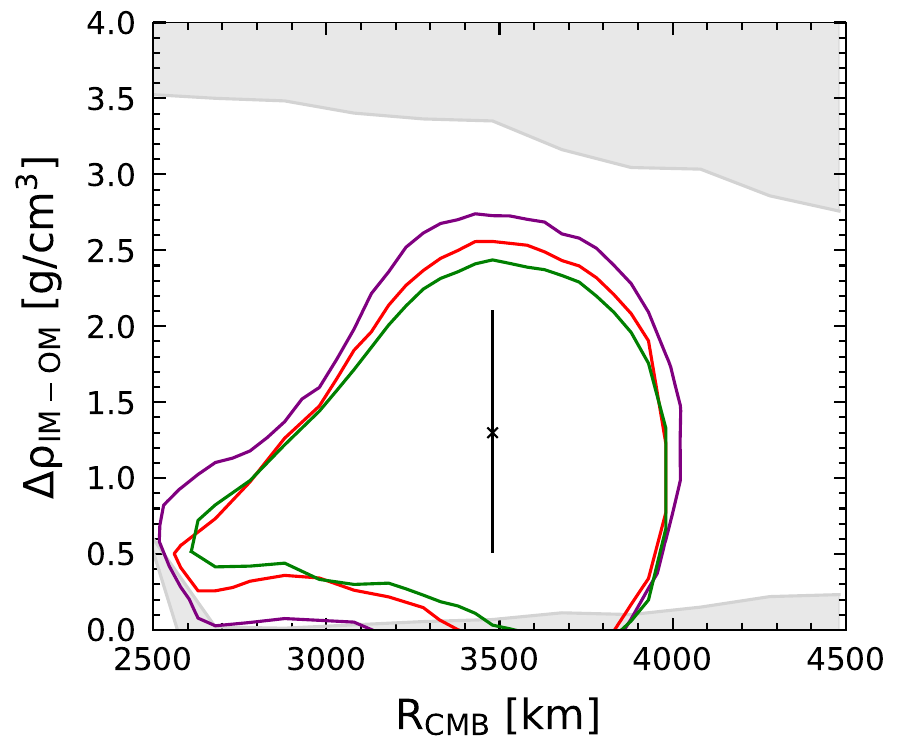}
	\includegraphics[width=0.45\linewidth]{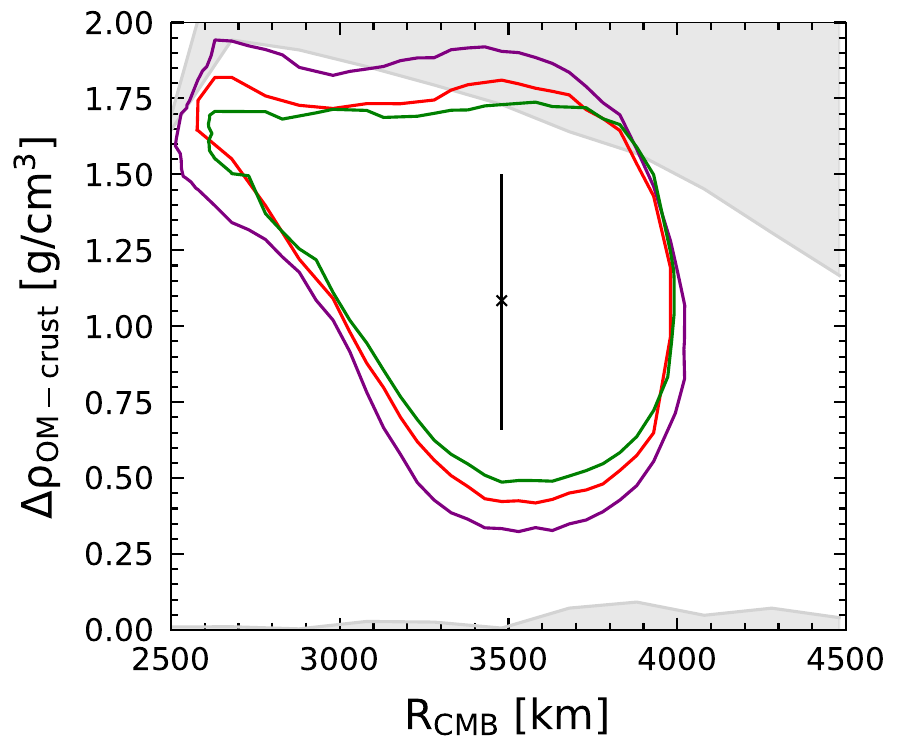}
	\mycaption{The median $\Delta\chi^2_\text{DJ-CMB}$ sensitivity contours in the plane of the density jumps and the location of the CMB radius at $1\sigma$ (2 d.o.f.) for three choices of the true values of $\sin^2\theta_{23}$. The violet, red, and green contours correspond to the sensitivities where MC data is generated using $\sin^2\theta_{23}$ (true) = 0.4, 0.5, and 0.6, respectively. The other details are the same as in Fig.~\ref{fig:2D_results}.}
	\label{fig:2D_results_impact_theta23}
\end{figure}

To analyze the impact of the true value of $\sin^2\theta_{23}$, we evaluate the expected sensitivity of the ICAL detector for the 1D-modification scenario in Fig.~\ref{fig:1D_results_impact_theta23}, and for the 2D-modification scenario in Fig.~\ref{fig:2D_results_impact_theta23}. In the fits, $\sin^2\theta_{23}$ is fixed at its corresponding true value, while all other oscillation parameters are fixed at their benchmark values, assuming NO. The remaining details in Fig.~\ref{fig:1D_results_impact_theta23} and Fig.~\ref{fig:2D_results_impact_theta23} follow those of Fig.~\ref{fig:1D_results} and Fig.~\ref{fig:2D_results}, respectively. We observe that the expected sensitivities of ICAL enhance (deteriorate) if the true $\theta_{23}$ belongs to the higher (lower) octant.

From both the Figs.~\ref{fig:1D_results_impact_theta23} and \ref{fig:2D_results_impact_theta23}, we observe that the expected ICAL sensitivities are larger (smaller) when $\theta_{23}$ lies in the higher (lower) octant.  This happens because, for NO, the dominant matter effect terms in $P(\nu_\mu \rightarrow \nu_\mu)$ survival probability and $P(\nu_e \rightarrow \nu_\mu)$ appearance probability are proportional to $\sin^2\theta_{23}$, as shown by the series expansion in Ref.~\cite{Akhmedov:2004ny}. With increasing $\sin^2\theta_{23}$, the survival probability $P(\nu_\mu \rightarrow \nu_\mu)$ decreases, whereas the appearance probability $P(\nu_e \rightarrow \nu_\mu)$ increases. Since the flux of $\nu_\mu$ is larger than that of $\nu_e$, the decrease in the contribution of the survival $(\nu_\mu \rightarrow \nu_\mu)$ channel is higher than the increase in that of the appearance $(\nu_e \rightarrow \nu_\mu)$ channel. Therefore, the resultant matter effect does not cancel out completely and is proportional to $\sin^2\theta_{23}$. This indicates that the ICAL detector would be able to constrain the density jump and the CMB location with higher statistical confidence if $\theta_{23}$ lies in the higher octant.

\section{Summary and conclusions}
\label{sec:conclusion}

The internal structure of Earth has been extensively studied through traditional methods such as gravitational and seismological measurements. In this paper, we utilize a completely different method to explore the internal structure of Earth, which relies on the weak interactions of neutrinos. This is broadly known as neutrino tomography of Earth. Using neutrinos as a non-traditional and complementary tool would pave the way for the ``multi-messenger tomography of Earth'', which would enhance our understanding of Earth's internal structure.  

A promising way of probing the internal structure of Earth is via matter effects in neutrino flavor transition. These matter effects depend upon the neutrino energy and the electron number density along their path, and alter the neutrino oscillation probabilities. The matter effects, particularly the PR/NOLR depend on the density jump at the CMB and its location, which in turn can alter neutrino oscillation probabilities.

An atmospheric neutrino detector can be sensitive to the value of the density jump at the CMB $(\Delta\rho_\text{CMB})$ and the CMB radius $(R_\text{CMB})$, by detecting the modified neutrino oscillation patterns. In the present work, we study the effect of modifying $\Delta\rho_\text{CMB}$ and $R_\text{CMB}$ on the neutrino oscillation probabilities and expected neutrino event distributions.

We quantify the expected sensitivity of the ICAL detector for determining the density jump at the CMB and its location for two kinds of modifications: (i) 1D-modifications of the density jump at the CMB with the fixed standard CMB location and (ii) 2D-modifications of the density jump at the CMB  as well as its location. We observe that, if the true mass ordering is normal, the ICAL detector would be able to measure the density jump at the standard CMB with a precision of about 15\% at $1\sigma$ with 1 Mt$\cdot$yr of exposure. The CID capability of the ICAL detector would play a significant role in achieving this precision; in the absence of CID, the $1\sigma$ sensitivity deteriorates to about 20\%. If the true mass ordering is inverted, the sensitivity is slightly worse. 

Exploiting the 2D-modifications, we present the $1\sigma$ sensitivity contours with and without the CID capability of the ICAL detector. Since the densities of all the layers modify simultaneously during the variation of the density jump at the CMB due to the imposed constraints, these results can also be interpreted in terms of the precisions on the density jumps at other layer boundaries, such as the IC-OC, IM-OM, and OM-crust. These sensitivities show that the neutrino oscillation data can further constrain the parameter space allowed by the measurements of $M_\text{E}$, $I_\text{E}$, the density $\rho_\text{crust}$, the radii $R_\text{IC}$, $R_\text{IM}$, $R_\text{OM}$, $R_\text{E}$, and the density ratio $\rho_\text{IC}/\rho_\text{OC}$. We also present the impact of the true value of $\sin^2\theta_{23}$ on the sensitivities of the ICAL detector to $R_\text{CMB}$ and density jumps. We observe that the sensitivities enhance (deteriorate) if the true value of $\theta_{23}$ belongs to the higher (lower) octant.

In the future, next-generation atmospheric neutrino detectors like Hyper-K, ORCA, DUNE, IceCube/DeepCore/Upgrade, and P-ONE with a large statistics of neutrino data will significantly enhance the precision related to the density profile of Earth using weak interactions of neutrinos. Even though many of these detectors do not have the CID capability, they may be able to compensate for it with a larger size or sensitivity to all neutrino flavors. The combined studies of the internal structure of Earth with the neutrino data will provide information that is complementary to that from the gravitational and seismological measurements. 

\subsection*{Acknowledgements} 
This study is performed by the INO-ICAL collaboration to explore the possibility of utilizing Earth's matter effects in oscillations of atmospheric neutrinos to constrain the core radius and density jumps inside Earth simultaneously. We would like to thank S. Goswami for her useful comments and constructive suggestions on our work. A.K.U. would like to thank the organizers of the ``International Workshop on Multi-messenger Tomography of the Earth (MMTE 2023)'' at APC-Universit$\acute{\text{e}}$ Paris Cit$\acute{\text{e}}$ in Paris, France, during 4th to 7th July 2023, for providing an opportunity to present the preliminary results from this work. We acknowledge the support of  the Department of Atomic Energy (DAE), Govt. of India, under the Project Identification Numbers RTI4002 and RIO 4001. S.K.A. acknowledges the financial support from the Swarnajayanti Fellowship (sanction order no. DST/SJF/PSA- 05/2019-20) provided by the Department of Science and Technology (DST), Govt. of India, and the Research Grant (sanction order no. SB/SJF/2020-21/21) provided by the Science and Engineering Research Board (SERB), Govt. of India, under the Swarnajayanti Fellowship project. A.K.U. acknowledges financial support from the DST, Govt. of India (sanction order no. DST/INSPIRE Fellowship/2019/IF190755). The numerical simulations are performed using the ``SAMKHYA: High-Performance Computing Facility'' at the Institute of Physics, Bhubaneswar, India. A.D. would like to acknowledge funding from the J. C. Bose Grant ANRF/JBG/2025/000265/PS of the Anusandhan National Research Foundation (ANRF), Government of India.

\begin{appendix}

\section{Role of CID capability in enhancing the sensitivity towards Earth's matter effects}
\label{app:sen_regions}

In this appendix, we aim to show the effective regions in the $(E^\text{rec}_\mu, \cos\theta^\text{rec}_\mu)$ plane that contribute to the sensitivity for the different analyses performed in this paper, with and without the CID capability.

\begin{figure}[htp!]
	\centering 
	\includegraphics[width=0.49\linewidth]{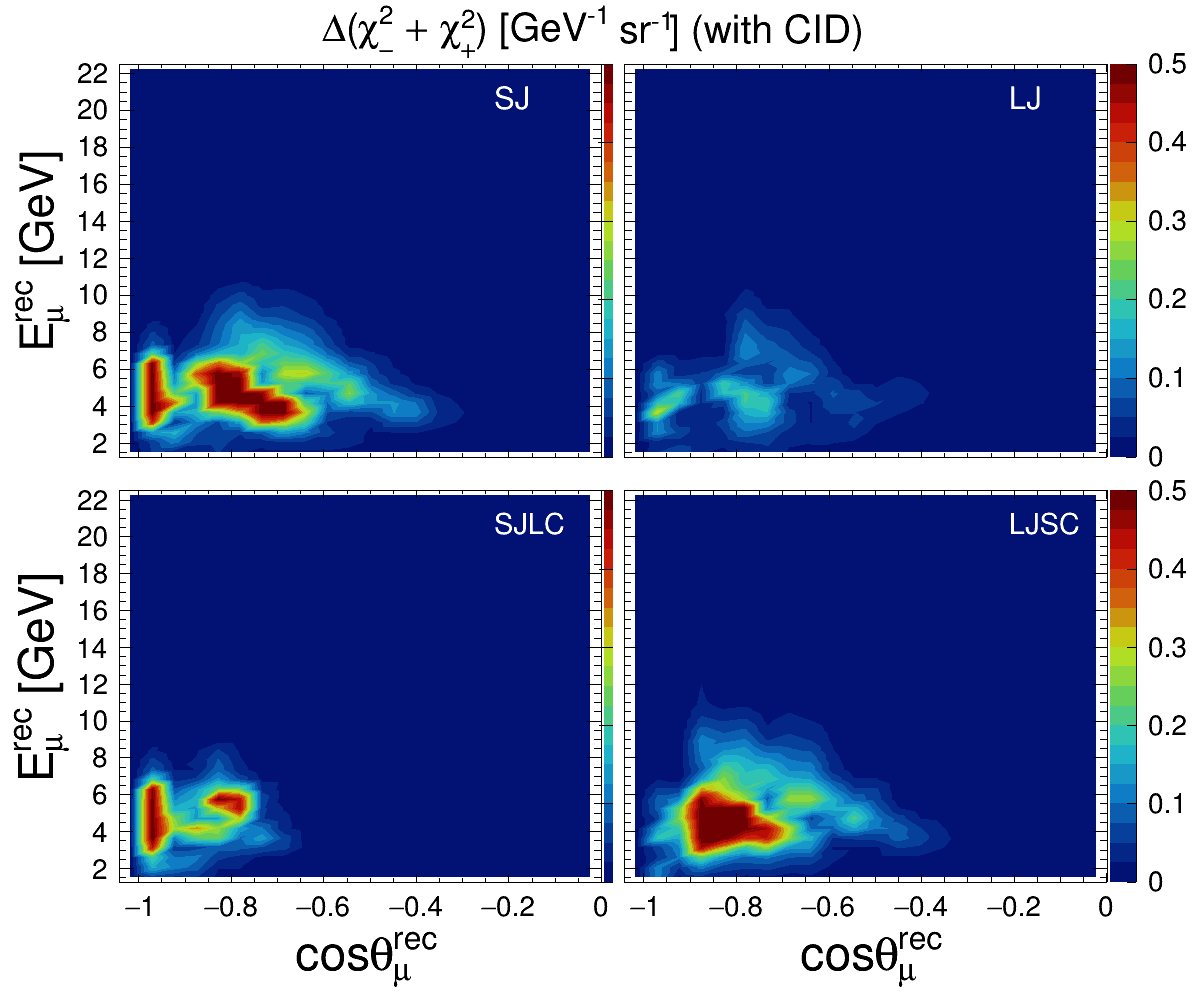}
	\includegraphics[width=0.49\linewidth]{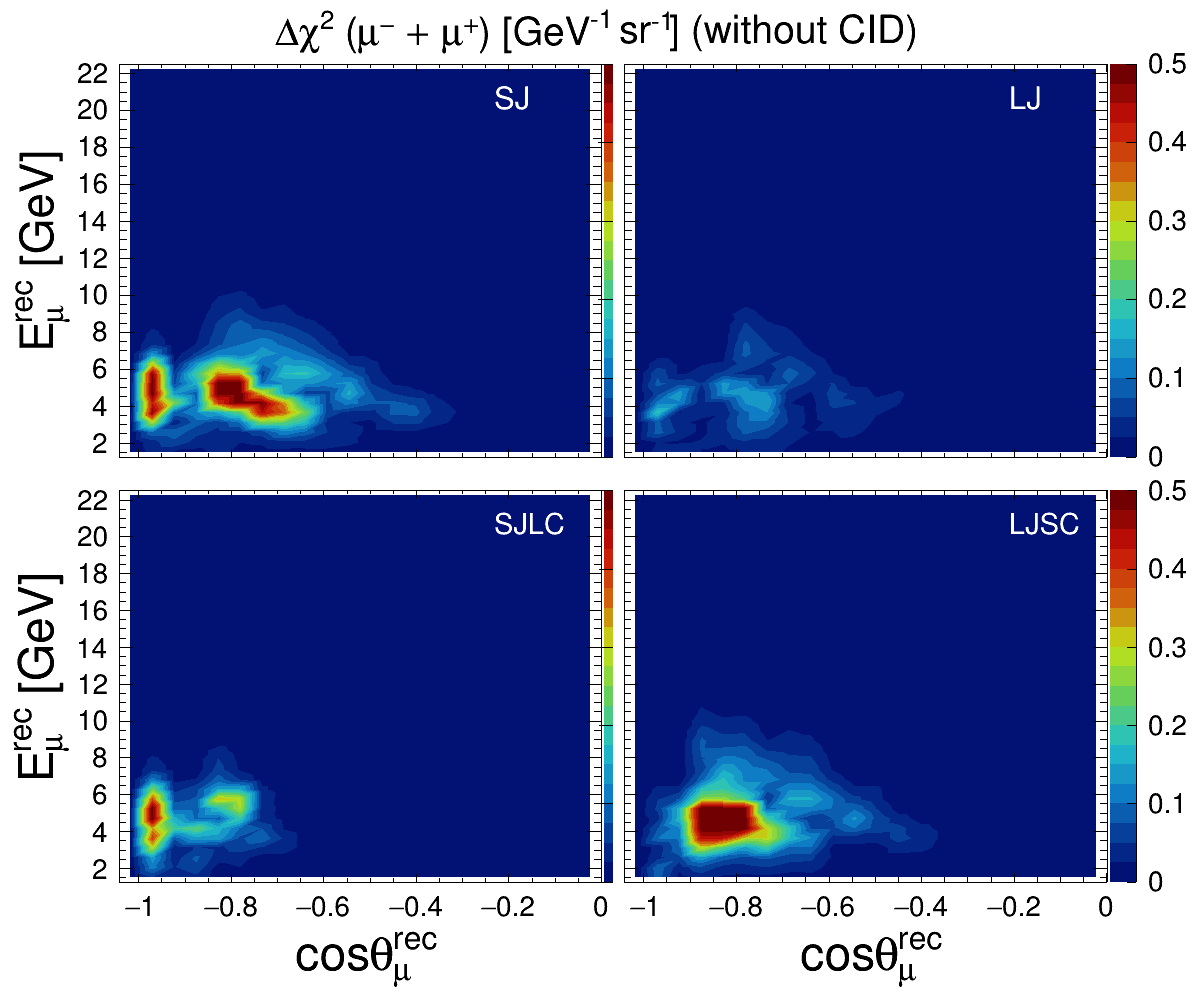}
	\caption{The distributions of $\Delta\chi^2$ (per unit area) in the $E^\text{rec}_\mu - \cos\theta^\text{rec}_\mu$ plane using 1 Mt$\cdot$yr of exposure of ICAL, assuming NO. The top (bottom) panels correspond to the SJ and LJ (SJLC and LJSC) modifications. The left (right) half shows the $\Delta\chi^2$ distributions with (without) CID capability.}
	\label{fig:chi2_distribution}
\end{figure}
	
In Fig.~\ref{fig:chi2_distribution}, we plot the distributions of $\Delta\chi^2$ [GeV$^{-1}$ sr$^{-1}$] in the plane of the reconstructed variables $E^\text{rec}_\mu$, and $\cos\theta^\text{rec}_\mu$. The top (bottom) panels show these distributions for the SJ and LJ (SJLC and LJSC) modifications as defined in Table~\ref{tab:density_jump_variation}. The plots in the left (right) half are obtained with (without) CID capability. While producing the plots without CID capability, we add the reconstructed $\mu^-$ and $\mu^+$ events in the same bin. In this figure, we add $\Delta\chi^2$ contributions from all four ${E^\prime}_{\text{had}}^\text{rec}$ bins for each $(E^\text{rec}_\mu,\cos\theta^\text{rec}_\mu)$ bin, and divide by $2\pi A$, where $A$ is the area of the $(E^\text{rec}_\mu,\cos\theta^\text{rec}_\mu)$ bin. For the sake of simplicity, we do not take into account the constant contribution to $\chi^2$ originating from the term involving the five pull parameters $\xi_l^2$ in Eq.~\ref{eq:chisq_mu-}, and we also do not minimize over the oscillation parameters in the fit.

This figure clearly demonstrates that the sensitivity to Earth's matter effects mainly arises from the bins which correspond to higher baselines $(\cos\theta^\text{rec}_\mu < - \, 0.6)$ and lower energies (2 GeV $< E^\text{rec}_\mu <$ 10 GeV), where the MSW and PR/NOLR resonances are prominent. Also, it is quite evident from this figure that the CID capability enhances the sensitivity in all four modifications. This enhancement is quite visible for the SJLC and LJSC scenarios (see the bottom panels).

\section{Sensitivity study for $R_\text{CMB}$ measurement}
\label{app:sens_rcmb_measurement}

In the 1D-modification discussed so far, we kept the $R_\text{CMB}$ fixed at 3480 km and only varied $\Delta\rho_\text{CMB}$. In this appendix, we present the sensitivity of the ICAL detector to measure the $R_\text{CMB}$ using both the three-layered and five-layered Earth density models. This comparison highlights the impact of moving from a three-layered to a five-layered density model on the determination of $R_\text{CMB}$.

We compare the sensitivity to determine the CMB radius for the following three scenarios:
\begin{itemize}
	\item {\bf Three-layered model with $M_\text{E}$ fixed:} For this scenario, we refer to the results of Case-II from our previous study~\cite{Upadhyay:2022jfd}, in which the core density is varied along with $R_\text{CMB}$ while keeping the Earth's total mass ($M_\text{E}$) fixed. In this scenario, the densities of the inner and outer mantle are kept constant.
	\item {\bf Five-layered model with $M_\text{E}$ fixed:} To ensure consistency with the previous study, we apply the same approach to the five-layered model, where we vary $R_\text{CMB}$ and the core density while keeping the Earth's total mass fixed. The densities of the mantle and crust remain unchanged.
	\item {\bf Five-layered model with $M_\text{E}$ and $I_\text{E}$ fixed:} Additionally, we also investigate a case in the five-layered model where both the Earth's total mass and its moment of inertia ($I_\text{E}$) are fixed while varying $R_\text{CMB}$. Under these constraints, the density of the inner mantle must also change in correlation with the core density.
\end{itemize}
In all three cases, the hydrostatic equilibrium condition is also imposed.

\begin{figure}[htp!]
	\centering
	\includegraphics[width=0.49\linewidth]{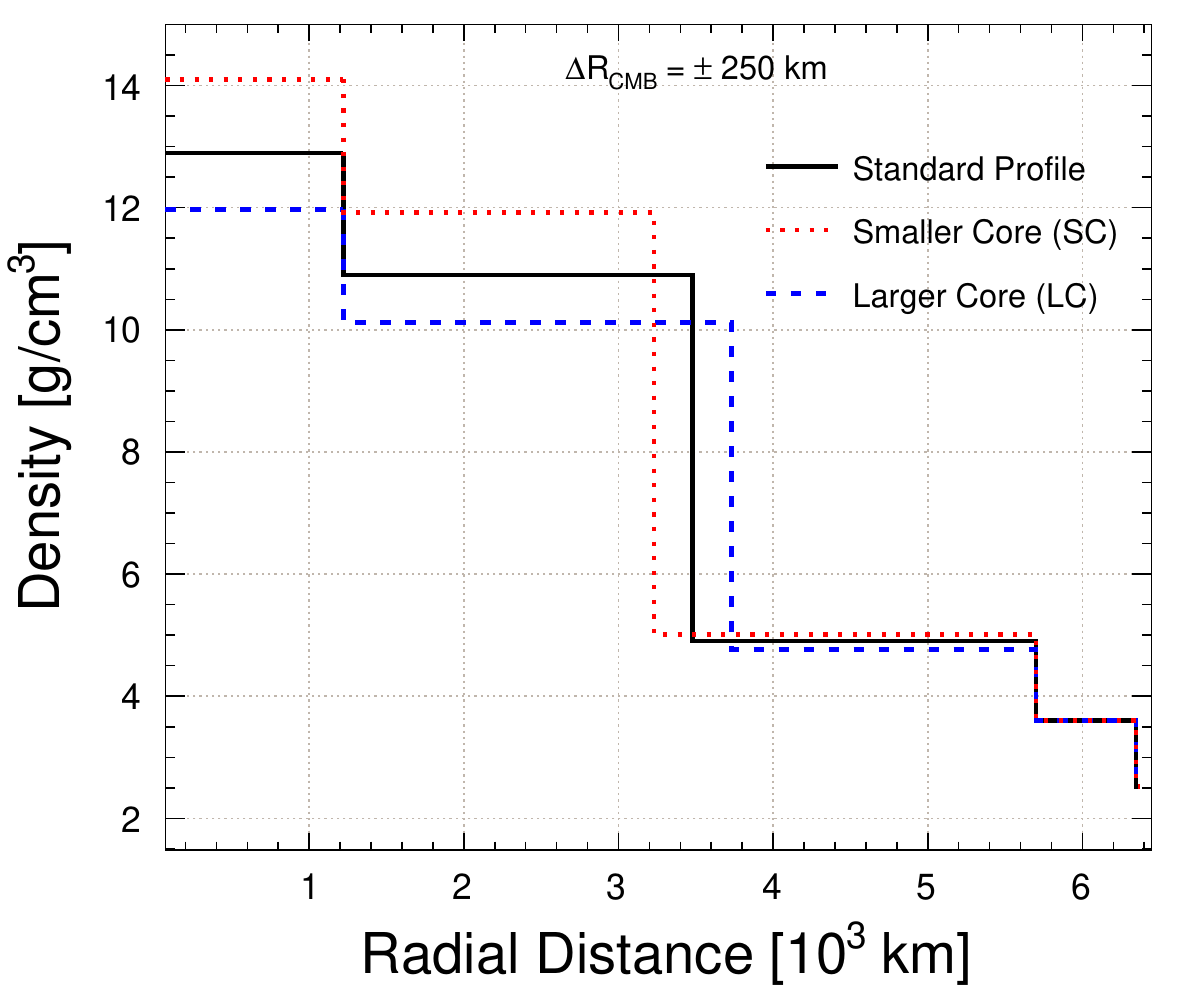}
	\includegraphics[width=0.49\linewidth]{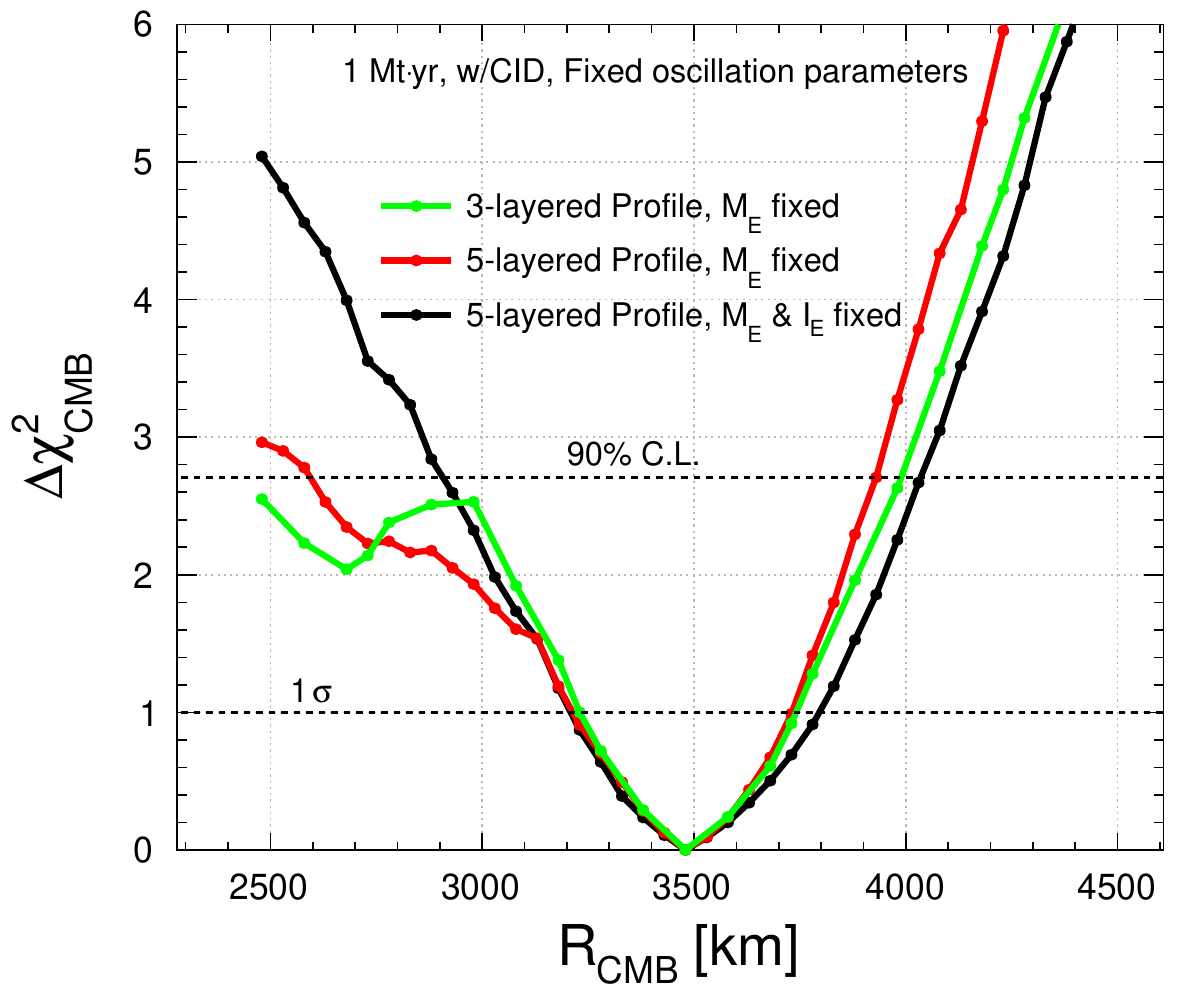} 
	\mycaption{The left panel shows the density profiles as functions of radial distance for modified values of $R_\text{CMB}$ in the five-layered PREM model. The black curve represents the standard PREM density profile with $R_\text{CMB} = 3480$ km. The dotted red (dashed blue) curve corresponds to the SC (LC) scenario, where $R_\text{CMB}$ is decreased (increased) by 250 km, while keeping the Earth's total mass and moment of inertia fixed. The right panel presents the median $\Delta\chi^2_\text{CMB}$ sensitivities as functions of $R_\text{CMB}$. The black curve corresponds to the five-layered PREM profile in which $R_\text{CMB}$ is varied under the constraints of fixed Earth's mass and moment of inertia (the same model as in the left panel). The green and red curves represent the sensitivities for (i) a three-layered profile and (ii) a five-layered profile, respectively, where $R_\text{CMB}$ is modified by constraining only the Earth's total mass. The green curve corresponds to Case-II in Ref.~\cite{Upadhyay:2022jfd}. All sensitivities are shown for an exposure of 1 Mt$\cdot$yr and include the CID capability of ICAL.}
	\label{fig:sens_rcmb_measurement}
\end{figure} 

For illustration, we show the modification in the CMB radius by $\Delta R_\text{CMB} = -\, 250$ km (smaller core or SC) and by $\Delta R_\text{CMB} = +\, 250$ km (larger core or LC) using a five-layered PREM profile in the left panel of Fig.~\ref{fig:sens_rcmb_measurement}. The black, dotted red, and dashed blue curves represent the density distributions as functions of radial distance for the standard $R_\text{CMB}$, the SC scenario, and the LC scenario, respectively. These modifications to the CMB location preserve both the Earth's mass and its moment of inertia.

The right panel of Fig.~\ref{fig:sens_rcmb_measurement} shows the median sensitivity of the ICAL detector in terms of $\Delta\chi^2_\text{CMB}$ as a function of $R_\text{CMB}$ for the scenarios described above. The black sensitivity curve is obtained by the five-layered PREM profile in which $R_\text{CMB}$ is varied while constraining both the Earth's mass and moment of inertia (consistent with the density profiles shown in the left panel). The green and red curves represent the sensitivities for (i) the three-layered profile and (ii) the five-layered profile, respectively, where $R_\text{CMB}$ is modified by constraining only the Earth's mass. The green curve corresponds to Case-II in Ref.~\cite{Upadhyay:2022jfd}. All oscillation parameters are kept fixed in the fit while evaluating $\Delta\chi^2_\text{CMB}$. The figure shows that the $1\sigma$ precision for determining $R_\text{CMB}$ remains $\pm \, 250$ km for the five-layered model when only the mass constraint is applied, similar to the three-layered case. When both the mass and moment of inertia constraints are included in the five-layered model, the precision degrades slightly to $\pm \, 290$ km. Overall, the results remain compatible with our previous study, and the transition to a more realistic five-layered density profile$-$while simultaneously applying all three global geophysical constraints (Earth's mass, moment of inertia, and hydrostatic equilibrium condition)$-$does not significantly affect the sensitivity to the CMB location. 

\end{appendix}



\bibliographystyle{JHEP}
\bibliography{References}

\end{document}